\documentclass[pdflatex,sn-mathphys-ay]{sn-jnl}

\usepackage[utf8]{inputenc}
\usepackage{graphicx}%
\usepackage{multirow}%
\usepackage{natbib}
\usepackage{pdflscape}
\usepackage{amsmath,amssymb,amsfonts}%
\usepackage{amsthm}%
\usepackage{mathrsfs}%
\usepackage[title]{appendix}%
\usepackage{xcolor}%
\usepackage{textcomp}%
\usepackage{manyfoot}%
\usepackage{booktabs}%
\usepackage{algorithmic}%
\usepackage{algorithm}%
\usepackage{listings}%
\usepackage{longtable}
\usepackage{mathtools}
\usepackage{graphicx}
\usepackage{multirow}
\usepackage{multicol}
\usepackage{algorithmic}
\usepackage{subfig}
\usepackage{todonotes}
\usepackage{algorithmic}
\usepackage{algorithm}
\usepackage{longtable}
\usepackage{hyperref}
\usepackage{orcidlink}
\usepackage{lmodern}

\graphicspath{{figures/}}

\begin{document}

\title[Fuzzy clustering and community detection: an integrated approach]{Fuzzy clustering and community detection: an integrated approach}

 \author[2,3]{\fnm{Domenico} \sur{Cangemi}}\email{dcangemi@luiss.it}
 \author[1]{\fnm{Pierpaolo} \sur{D'Urso}\orcidlink{0000-0002-7406-6411}}\email{pierpaolo.durso@uniroma1.it}
 \author[2,3]{\fnm{Livia} \sur{De Giovanni}\orcidlink{0000-0001-7806-2170}}\email{ldegiovanni@luiss.it}
 \author[2,3]{\fnm{Lorenzo} \sur{Federico}\orcidlink{0000-0003-3231-3901}}\email{lfederico@luiss.it}
 \author[1]{\fnm{Vincenzina} \sur{Vitale}\orcidlink{0000-0001-9512-3609}}\email{vincenzina.vitale@uniroma1.it}

 \affil[1]{\orgdiv{Department of Social Science and Economics}, \orgname{Sapienza University}, \orgaddress{\street{Piazzale Aldo Moro 5}, \city{Rome}, \postcode{00185}, \state{Lazio}, \country{Italy}}}

 \affil[2]{\orgdiv{Department of Political Science}, \orgname{Luiss University}, \orgaddress{\street{Viale Romania 32}, \city{Rome}, \postcode{00197}, \state{Lazio}, \country{Italy}}}

 \affil[3]{\orgdiv{Data Lab}, \orgname{Luiss University}, \orgaddress{\street{Viale Pola 12}, \city{Rome}, \postcode{00198}, \state{Lazio}, \country{Italy}}}


\abstract{
This paper addresses the ambitious goal of merging two different approaches to group detection in complex domains: one based on fuzzy clustering and the other on community detection theory. To achieve this, two clustering algorithms are proposed: \textit{Fuzzy C-Medoids Clustering with Modularity Spatial Correction} and \textit{Fuzzy C-Modes Clustering with Modularity Spatial Correction}. The former is designed for quantitative data, while the latter is intended for qualitative data. 
The concept of fuzzy modularity is introduced into the standard objective function of fuzzy clustering algorithms as a spatial regularization term, whose contribution to the clustering criterion based on attributes is controlled by an exogenous parameter.
An extensive simulation study is conducted to support the theoretical framework, complemented by two applications to real-world data related to the theme of sustainability. The first application involves data from the 2030 Agenda for Sustainable Development, while the second focuses on urban green spaces in Italian provincial capitals and metropolitan cities.
Both the simulation results and the applications demonstrate the advantages of this new methodological proposal.
}




\captionsetup[figure]{labelfont={bf},labelformat={default},labelsep=colon,name={Fig.}}
\captionsetup[table]{labelfont={bf},labelformat={default},labelsep=colon,name={Table}}

\maketitle

\section{Introduction}
This study aims to merge two methodologies — \textit{fuzzy clustering} and \textit{community detection} — to uncover natural groupings of objects. Both approaches tackle the same underlying problem, \textit{i.e} the unsupervised classification of objects, but from distinct perspectives. Fuzzy clustering centers on the attributes of the objects, allowing for partial membership to multiple clusters, while community detection looks at the connections between nodes (or objects) within a network, aiming to group nodes based on their connectivity patterns.

By bridging these two approaches, the study aims to leverage the strengths of both techniques to better understand and detect underlying structures in complex datasets.  This integrated methodology is highly applicable across various domains, including social networks, biological systems, and economic and social data.

From this perspective, the $N$ objects to be partitioned serve a dual role: they are both the statistical units on which attributes are measured and the nodes within a network. This dual perspective enables the clustering process to integrate attribute-based similarities and network adjacency, resulting in partitions that capture not only the inherent characteristics of the objects but also their connections within the network structure.  This innovative approach builds on the seminal work of \citet{d2024fuzzy}, which allows the joint clustering of two sets of statistical units by leveraging two sets of attributes (one for each group of units) along with the bi-adjacency matrix derived from a bipartite graph.

In the present paper, in particular, two novel clustering techniques are  proposed, the  \textit{Fuzzy C-medoids clustering with modularity spatial correction} and the \textit{Fuzzy C-Modes clustering with modularity spatial correction}. The former is here proposed for quantitative variables, while the latter is for qualitative data.
The term "spatial correction" is justified by the fact that, as explained later in this paper, we incorporate information about the graph structure into the traditional objective function — usually based solely on attributes — as a regularization term. The influence of this regularization term on the clustering criterion is adjusted through an exogenous tuning parameter.

From a methodological point of view, both techniques uses the fuzzy entropy approach as early proposed by  \citet{miyamoto2, li1999gaussian} and \citet{miyamoto1} and then extensively applied \citep{yao2000entropy, ichihashi2000gaussian, zarinbal2014relative, kahali2019new, gao2019novel}. Another interesting extension to time-varying data can be found in \cite{coppi}.

Specifically, recent proposals that combine the fuzzy entropy and the medoids-based approach can be found in \cite{d2022kemeny}, \cite{dursofuzzy}, \cite{spatial}, \cite{drobust}, \cite{tail}, \cite{vitale2024entropy}.

In this regard, we remember that Fuzzy Partitioning Around Medoids (PAM) approach \citep{krishnapuram1, krishnapuram2}  offers several key advantages, the foremost being that each cluster is represented by a prototype called "medoid" that, unlike centroids which can be abstract and may not correspond to actual data points, is a real object from the dataset itself. This makes the medoid-based representation more interpretable and practical.

Furthermore, it is worth noticing that the concept of centroid as cluster prototype becomes even more unfeasible when the set of possible values for the attributes does not follow a vector space structure — such as in the case of categorical data. 

A widely recognized method for clustering categorical data is the C-modes algorithm \citep{Huang1997AFC, chaturvedi2001kmodes}. This algorithm seeks to minimize the dissimilarity between each object and the mode of its assigned cluster, where the mode is defined as the vector of the most frequent values for each categorical attribute within the cluster. Dissimilarity is typically quantified using the simple matching dissimilarity measure, which counts the number of mismatched attributes between two objects.

A well-known extension to the fuzzy framework has been proposed by \cite{huang1999fuzzy} and, then, modifications to the simple matching dissimilarity measure were introduced within the fuzzy C-modes algorithm by \cite{NG2009}, enhancing its ability to create clusters with strong intra-cluster similarity. Other interesting variants can be found in \cite{Gan, KIM20041263, SAHA2015422, Bai2014,LIU2020106425}. 

However, to the best of our knowledge, no versions of fuzzy C-medoids or fuzzy C-modes algorithms exist that have already integrated the modularity concept in the objective function. 
More specifically, we propose this original contribution starting from the idea that modularity, designed by \cite{newman2006modularity} to optimize community structures in networks, can ensure that clusters based on attributes correspond more naturally to real-world communities or subgroups. The modularity of a partition of the vertex set of a network is the difference between the observed number of connections within the sets of the partition minus its expected value in a suitable null model so that a partition with high modularity has more connections within its sets than expected. Exact modularity-maximization is NP-hard \citep{brandes2007modularity}, so approximate algorithms are employed \citep{ClaNew04,traag2019louvain}. Further, for networks with specific properties, such as a multipartite structure, modified version of the modularity, with different null models, have been designed \citep{barber2007modularity, neubauer2011tripartite}.
The combination of two techniques from the different fields of community detection and fuzzy clustering inherits all the advantages and properties from both classes of models, thereby enhancing flexibility.
\\
The outline of the paper is as follows. Section 2 first defines fuzzy modularity and then provides the mathematical details of the two clustering techniques. After defining the proposed internal validity measure in Section 3, Section 4 presents and discusses an extensive simulation plan. Section 5 focuses on the application to two real-world datasets. Last section concludes with some remarks and considerations.

\section{Fuzzy clustering models with community detection}

In this section, we recall the notion of Fuzzy Modularity of a partition of the vertices of a network, originally introduced in \cite{nepusz2008fuzzy} as a way to find "bridge" nodes in communities in networks, that is, nodes that are connected to other nodes in multiple communities. This is the fuzzy equivalent of the original "crisp" modularity defined by \cite{newman2006modularity}. Here we use it for a completely different purpose, that is, to cluster together units that have both an adjacency structure represented as a network, and a set of attributes which can take various types of values (quantitative, ordinal, qualitative, etc...).

In practice, we require our dataset to be structured as follows:

\begin{itemize}
    \item A matrix of attributes $\mathbf{X}:=\{x_{n,i}, n\leq N,i\leq I\}$, where each row represents one of the $N$ units and each column one of the $I$ attributes. The entry $x_{n,i}$ indicates the value of attribute $i$ observed on unit $n$. 
   \item An adjacency matrix $\mathbf{A}:=\{a_{n,m}, n,m\leq N\}$, which can both be a binary matrix where $a_{n,m}=1$ if $n$ and $m$ are adjacent and $a_{n,m}=0$ otherwise, or a non-negative matrix where $a_{n,m}\geq 0$ represents the strength of the connection between $n$ and $m$. In this paper we will focus on the first case.
\end{itemize}
Note that in principle $x_{n,i}$ can be itself a more complex data structure than just a single value, like a sequence of values indicating a sample or a time series, as long as it is possible to define a suitable dissimilarity function $d(x,y)$ between any two points $x, y$ in the attribute space. In the present paper we will consider the cases where $x_{n,i}$ is either a real number or an attribute taking values in an unordered space.

Moreover, the matrix $\mathbf{A}$ can represent either contiguity in a physical space or a connection in a more abstract sense (friendship between people, trade relations between countries...) and in Sections \ref{sec:AppMd} and \ref{sec:appMo} we will present an example of each.  

We want to develop an algorithm that can capture together the properties of the attributes and of the network and output a fuzzy partition that has both these properties:

\begin{itemize}
    \item \textbf{Cluster Cohesion}: If two units $n_1,n_2 \leq N$ have high membership to the same cluster $c$, the dissimilarity of their attributes is likely to be small.
    \item \textbf{Community Validity }: Couples of units $n_1, n_2 \leq N$ which both have high membership to the same cluster $c$, are more likely to be adjacent than the average.
\end{itemize}

The partition will be represented as a matrix $\mathbf{U}:=\{u_{n,c}, n \leq N, c\leq C\}$ where $u_{n,c}$ represents the degree of membership of unit $n$ to cluster $c$. These memberships are constrained so that  
\begin{equation}\label{eq:constraints}
     u_{n,c} \geq 0, \quad \sum_{c=1}^{C}u_{n,c}=1.
\end{equation}
To produce the partition $\mathbf{U}$, we optimize an objective function that is the linear combination of the objective function of a fuzzy entropic clustering algorithm and the partition's fuzzy modularity. We will introduce in detail these concepts in the remaining part of this section.

\subsection{Fuzzy Modularity}

We lift the definition of the fuzzy modularity from \cite{nepusz2008fuzzy}. We define for every unit $n$ its \emph{strength} in the network represented by $\mathbf{A}$ as $w_{n}=\sum_{m=1}^N a_{n,m}$, and the total strength of the network as $L=\sum_{n=1}^N w_{n}$. Note that in the case of a binary unweighted network, $w_n$ is the degree of node $n$ and $L$ is twice the total number of edges in the network.

We define the modularity matrix $\mathbf{B}:=\{b_{n,m}, n,m\leq N\}$ as

\begin{equation}
    b_{n,m}= a_{n,m}- \frac{w_nw_n}{L}
\end{equation}

We can thus define, up to a constant, the fuzzy modularity of the partition $\mathbf U$ with respect to the network $\mathbf{A}$ as 

\begin{equation}\label{eq:uCond}
    Q (\mathbf U,\mathbf{A})=\sum_{n=1}^N\sum_{m=1}^Nb_{n,m}\sum_{c=1}^C u_{n,c}u_{m,c}(1-\delta_{n,m}).
\end{equation}

Here, and in the rest of the paper, $\delta$ is the Kronecker $\delta$, the indicator that $n$ and $m$ are equal.
As mentioned in \cite{nepusz2008fuzzy}, this is equivalent to the expected value of the crisp modularity of a random partition where each unit $n$ is assigned to cluster $c$ with probability $u_c$ independently of all the other.

Note that $b_{n,m}<0$ indicates that the connection between $n$ and $m$ is absent or is weaker than expected, and $b_{n,m}>0$ means instead that the connection is stronger than expected, and that $\sum_{c=1}^C u_{n,c}u_{m,c}$ is high when $n$ and $m$ have high membership to the same cluster. This means that a high value of $Q$ indicates that units with high membership to the same cluster have typically stronger connections than usual.

We present two models which use modularity-based spatial regularization to showcase its effectiveness, the Fuzzy C-Medoids with modularity spatial correction (FCMd-MSC) and the Fuzzy C-Modes with modularity spatial correction (FCMo-MSC). 

\subsection{ Fuzzy C-Medoids with modularity spatial correction}

We start by introducing the FCMd-MSC, here we assume that the attributes are defined in a metric space, so that for any two units is defined the distance $d(\mathbf{x}_{n},\mathbf{x}_{m})$ between the respective attributes. Our goal is to produce, besides the partition matrix  $\mathbf{U} $, also a set of prototype units, the medoids, $(\mathbf{x}_1,\ldots,\mathbf{x}_c,\ldots,\mathbf{x}_C)$, one for each cluster, as introduced in \cite{krishnapuram1999fuzzy}. To do so, we minimize an objective function of the fuzzy partition $\mathbf{U}$ and the medoids, written in terms of the attribute matrix $\mathbf{X}$ and the adjacency matrix $\mathbf{A}$. To balance the information provided by the attributes and by the network structure, we optimize a convex combination of the objective function of a fuzzy entropic C-medoids algorithm on the matrix $\mathbf{X}$ and the fuzzy  modularity of the network defined by  the adjacency matrix $\mathbf{A}$. This function is further defined by the parameters $\gamma$, which controls the relative importance of the two matrices, $C$ which corresponds to the number of clusters and $p$, which tunes the fuzziness of the partition.
We thus attempt to solve the following minimization problem, constrained by the conditions in \eqref{eq:constraints}

\begin{align}\label{eq:objectiveMd}\nonumber
\min_{\mathbf{U}, \mathbf{x}_c}J_{p,C,\gamma}(\mathbf{U}, \mathbf{x}_c) :=(1-\gamma)\sum\limits_{n=1}^{N}&\sum\limits_{c=1}^{C}u_{n,c}d^2(\mathbf{x}_{n},\mathbf{x}_{c})+p\sum_{n=1}^N\sum\limits_{c=1}^{C}u_{n,c}\log (u_{n,c})\\&-\frac{\gamma}{2} \sum\limits_{n=1}^{N}\sum\limits_{c=1}^{C}\sum\limits_{m=1}^N u_{n,c}b_{n,m}u_{m,c} (1-\delta_{n,m}).
\end{align}

where $d^2(\mathbf{x}_{n},\mathbf{x}_{c})$ is the squared Euclidean distance between the $n$-th object and the $c$-th medoid.

Using the Lagrangian multiplier method (see proof in sect.\ref{app:lagrangian}) we find that the minima with respect to the memberships $u_{i,c}$ of $J_{p,C,\gamma}(\mathbf{U}, \mathbf{x}_c)$ are obtained as

\begin{equation}
		u_{nc}=\frac{	    	
			\exp\left\{-\frac{1}{p}\left(
			(1-\gamma)d^2(\mathbf{x}_{n},\mathbf{x}_{c})-
			\gamma \sum\limits_{{m= 1}}^N b_{n,m}u_{m,c} (1-\delta_{n,m})
			\right)\right\}
		}
		{
		\sum\limits_{c'=1}^C \exp\left\{-\frac{1}{p}\left((1-\gamma)
			d^2(\mathbf{x}_{n},\mathbf{x}_{c'})-
			\gamma \sum\limits_{{m= 1}}^N b_{n,m}u_{m,c'} (1-\delta_{n,m})
			\right)\right\}.
		}
\label{eqn:UMd}
\end{equation}
The optimization over the choice of the medoids $(\mathbf{x}_1,\ldots,\mathbf{x}_c,\ldots,\mathbf{x}_C)$ is done by brute-force search.

We summarize the optimization procedure in Algorithm \ref{alg:md}.

	\begin{algorithm}[htbp]
		\caption{Fuzzy C-Medoids with modularity spatial correction (FCMd-MSC)}\label{alg:md}
		\begin{algorithmic}[1]
			\STATE Fix $C$, $max.iter$, $conv=1 \times 10^{-9}$ and initialize randomly the membership degree matrix $\mathbf U$;
			\STATE Set $iter=0$;			
			\STATE Set \emph{medoids}:= $ ( \mathbf{x}_{1},\ldots,\mathbf{x}_{C})$, arbitrarily;
			\REPEAT 
			\STATE Set $\mathbf{U_{old}}=\mathbf{U}$;
                \STATE Update \emph{medoids} as follows: 
			\FOR{$c=1 \text{ to } C$}
                \STATE Define \emph{members}=$\{i\leq N : c= \arg\max_{1\leq k \leq C} u_{i,k}\}$
                \IF {\emph{members} is not empty}
			\STATE $q=\arg\min_{q \in members} \sum\limits_{n=1}^{N}u_{n,c}d^2(\mathbf{x}_{n},\mathbf{x}_{q})$ 
			\STATE Set $\Rightarrow \mathbf{x}_{c}=\mathbf{x}_{q}$
                \ENDIF
			\ENDFOR

            \STATE Update $\mathbf{U}$ using (\ref{eqn:UMd});

			\STATE $iter \gets iter+1$;
			\UNTIL $\lVert \mathbf{U_{old}}-\mathbf{U}\rVert_1 < conv$ or $iter=max.iter$
            \RETURN $\mathbf U$, $ ( \mathbf{x}_{1},\ldots,\mathbf{x}_{C})$.
		\end{algorithmic}
	\end{algorithm}
\subsubsection{Calculation of the derivatives of the Lagrangian function}
\label{app:lagrangian}
In this section we show in detail the computations to derive equation  \eqref{eqn:UMd}  which is necessary in Algorithm \ref{alg:md} to optimize the membership matrix $\mathbf{U}$ at each iteration.

We write the Lagrangian function for every $u_{n,c}$:

\begin{align}\label{eq:LagMd}
L_{Md}(u_{n,c},\lambda) =&(1-\gamma)\sum\limits_{n=1}^{N}\sum\limits_{c=1}^{C}u_{n,c}d^2(\mathbf{x}_{n},\mathbf{x}_{c})+p\sum_{n=1}^N\sum\limits_{c=1}^{C}u_{n,c}\log (u_{n,c})\\&-\frac{\gamma}{2} \sum\limits_{n=1}^{N}\sum\limits_{c=1}^{C}\sum\limits_{m=1}^N u_{n,c}b_{n,m}u_{m,c} (1-\delta_{n,m})- \lambda \Big(\sum_{c=1}^{C}u_{n,c} -1\Big).\nonumber
\end{align}

From this we compute first the derivative with respect to $\lambda$

\begin{equation}
    \frac{\partial L_{Md}(u_{n,c},\lambda)  }{\partial \lambda}= \sum_{c=1}^{C}u_{n,c} -1,
\end{equation}
so that
\begin{equation}\label{eq:nlambda0}
    \frac{\partial L_{Md}(u_{n,c},\lambda)  }{\partial \lambda}=0 \Longleftrightarrow \sum_{c=1}^{C}u_{n,c} =1.
\end{equation}
And then the derivative with respect to $u_{n,c}$

\begin{equation}
     \frac{\partial L_{Md}(u_{n,c},\lambda)  }{\partial u_{n,c}}=(1-\gamma)d^2(\mathbf{x}_{n},\mathbf{x}_{c})+p(\log u_{n,c}+1)-\gamma \sum_{m=1}^N b_{n,m}u_{m,c} (1-\delta_{n,m}) - \lambda
\end{equation}
We then solve in $u_{n,c}$:

\begin{align}
    \log u_{n,c} = &\frac{\lambda}{p} -\frac{(1-\gamma)}{p}d^2(\mathbf{x}_{n},\mathbf{x}_{c})+\frac{\gamma}{p}\sum_{m=1}^{N}b_{n,m}u_{m,c} (1-\delta_{n,m})-1 \Longleftrightarrow \\
    \frac{\partial L_{Md}(u_{n,c},\lambda)  }{\partial u_{n,c}}=&0  \qquad \forall c \leq C \quad n\leq N.\nonumber
\end{align}

Consequently, taking the exponential of both terms,

\begin{align}\label{eq:uexp}
    u_{n,c}=\exp\Big\{\frac{\lambda}{p} -\frac{(1-\gamma)}{p}d^2(\mathbf{x}_{n},\mathbf{x}_{c})+\frac{\gamma}{p}\sum_{m=1}^{N}b_{n,m}u_{m,c} (1-\delta_{n,m})-1\Big\}
\end{align}

Recalling \eqref{eq:nlambda0} we find that

\begin{align}
    \exp\left\{ \frac{\lambda}{p}-1\right\}=\frac{1}{\sum_{c=1}^{C}\exp\left\{ -\frac{(1-\gamma)}{p}d^2(\mathbf{x}_{n},\mathbf{x}_{c})+\frac{\gamma}{p}\sum_{m=1}^{N}b_{n,m}u_{m,c} (1-\delta_{n,m})\right\}},
\end{align}
so that, substituting back in \eqref{eq:uexp} we obtain \eqref{eqn:UMd}.

\subsection{ Fuzzy C-Modes with modularity spatial correction}

The other algorithm we introduce is the FCMo-MSC, where each attribute $i$ takes value in a discrete unordered set $\Omega_i$. Here we define the distance between two elements in the space of vectors of attributes  as the simple matching, or Hamming, distance, that is

\begin{equation}
    d_{SM}(\mathbf{x},\mathbf{y})=\sum_{i=1}^I (1-\delta_{x_i,y_i}) \quad \forall  \mathbf{x},\mathbf{y} \in \bigtimes_{i=1}^I \Omega_i.
\label{eqn:HamD}
\end{equation}
Here, instead of representing each cluster by its more "central" unit, we represent it by the vector $\mathbf{\hat x}_c$ of modes  of each attribute over the cluster, that is 

\begin{align}\label{eq:moddef}
\hat x_{i,c} = \arg\max_{\Omega_i} \sum_{n=1}^N u_{n,c}\delta_{\hat{x}_{c,i},x_{n,i}}.
\end{align}

The clusters are obtained by solving the minimization of the following objective function, which is a convex combination of the fuzzy modularity of the partition $\mathbf{U}$ over a network represented by the adjacency matrix $\mathbf{A}$ and the objective function of a fuzzy entropic $C$-modes over the attribute matrix $\mathbf{X}$ subject to \eqref{eq:constraints}. We thus have to solve the following optimization problem:

\begin{align}\label{eq:objectiveMo}\nonumber
\min_{\mathbf{U}, \mathbf{\hat x}_c}F_{p,C,\gamma}(\mathbf{U}, \mathbf{\hat x}_c) :=(1-\gamma)\sum\limits_{n=1}^{N}&\sum\limits_{c=1}^{C}u_{n,c}d_{SM}^2(\mathbf{x}_{n},\mathbf{x}_{c})+p \sum_{n=1}^N\sum\limits_{c=1}^{C}u_{n,c}\log (u_{n,c})\\&-\frac{\gamma}{2} \sum\limits_{n=1}^{N}\sum\limits_{c=1}^{C}\sum\limits_{m=1}^N u_{n,c}b_{n,m}u_{m,c} (1-\delta_{n,m}).
\end{align}
The solution is approximated optimizing iteratively over the membership and over the modes. The optimum for the choice of the modes for a given membership matrix $\mathbf{U}$ is given in \eqref{eq:moddef}, while the optimum for given modes over the membership is found using the Lagrangian multipliers method \footnote{Following proof in sect.\ref{app:lagrangian}, the derivation of equation \eqref{eqn:UMo} is identical except for the fact that the Euclidean distance is replaced by the simple matching distance} as follows:
\begin{equation}
		u_{nc}=\frac{	    	
			\exp\left\{-\frac{1}{p}\left(
			d_{SM}^2(\mathbf{x}_{n},\mathbf{\hat x}_{c})-
			\gamma \sum\limits_{{m= 1}}^N b_{n,m}u_{m,c} (1-\delta_{n,m})
			\right)\right\}
		}
		{
		\sum\limits_{c'=1}^C \exp\left\{-\frac{1}{p}\left(
			d_{SM}^2(\mathbf{x}_{n},\mathbf{\hat x}_{c'})-
			\gamma \sum\limits_{{m= 1}}^N b_{n,m}u_{m,c'} (1-\delta_{n,m})
			\right)\right\}
		}
\label{eqn:UMo}.
\end{equation}

We summarize the optimization procedure in Algorithm \ref{alg:mo}.

	\begin{algorithm}[htbp]
		\caption{Fuzzy C-Modes with modularity spatial correction (FCMo-MSC)}\label{alg:mo}
		\begin{algorithmic}[1]
			\STATE Fix $C$, $max.iter$, $conv=1 \times 10^{-9}$ and initialize randomly the membership degree matrix $\mathbf U$;
			\STATE Set $iter=0$;			
			\STATE Set \emph{modes}:= $ ( \mathbf{\hat x}_{1},\ldots,\mathbf{\hat x}_{C})$, arbitrarily;
			\REPEAT 
			\STATE Set $\mathbf{U_{old}}=\mathbf{U}$;
                \STATE Update \emph{modes} as follows: 
			\FOR{$c=1 \text{ to } C$}
            \FOR{$i=1 \text{ to } I$}
			\STATE Set $\hat x_{i,c} = \arg\max_{\Omega_i} \sum_{n=1}^N u_{n,c}\delta_{\hat{x}_{c,i},x_{n,i}}$.
			\ENDFOR
            \ENDFOR
            \STATE Update $\mathbf{U}$ using (\ref{eqn:UMo});

			\STATE $iter \gets iter+1$;
			\UNTIL $\lVert \mathbf{U_{old}}-\mathbf{U}\rVert_1 < conv$ or $iter=max.iter$
            \RETURN $\mathbf U$, $ ( \mathbf{\hat x}_{1},\ldots,\mathbf{\hat x}_{C})$.
		\end{algorithmic}
	\end{algorithm}

\subsection{The rationale behind modularity correction}\label{subsec:rationale}

In this section, we discuss the reasons to use fuzzy modularity as a regularization term and the potential advantages it has compared to other spatial corrections. We will not focus on techniques based on a metric or model-based approach (see e.g. \cite{disegna2017copula, d2020robust, d2022community}), as this approach works on a completely different basis and has different requirements on the spatial structure of the data, for example, it requires them to exist in a broader metric space. This, is not always the case, as we will show in Section \ref{sec:AppMd}, where we will apply Algorithm \ref{alg:md} on an adjacency network among nations based on international agreements and not geographical proximity. We will instead compare it to network-based spatial regularization, where, like in our case, the geometric nature of the data is encoded by the adjacency matrix of a network.

In many spatial models that are based on an adjacency matrix, such as that introduced in \cite{Pham2001} and \citet{d2019spatial_temporal,d2022covid,DURSO2025100874,lopez2021spatial}, the spatial term is treated as an addition to the clustering objective function that is meant to influence the membership of the units in the fuzzy clustering, but is not a clustering objective function on its own. Indeed, the spatial penalty has the form 

\begin{equation}\label{eq:spaOld}
    \frac{\beta}{2}\sum\limits_{n=1}^{N}\sum\limits_{c=1}^{C}u_{n,c}
	\sum\limits_{m=1}^{M}\sum\limits_{{c'\in C_c}}a_{n,m}u_{m,c'},
\end{equation}
where $C_c$ is the set of all clusters except $c$.
Given that there is a penalty for putting in separate clusters units that are adjacent, but not for putting in the same cluster units that are not adjacent, it is easy to verify that the partition that minimizes the value in \eqref{eq:spaOld} is one where all units have membership $1$ to the same cluster and $0$ to all the other clusters.

Instead, the modularity spatial correction has the goal to produce a partition that in some sense ''interpolates" between the partition that maximizes the agreement with respect to the attributes (ideally obtained setting $\gamma=0$) and that that maximizes the agreement with respect to the adjacency network (ideally obtained setting $\gamma=1$). This second partition now is in principle not a trivial one. 

We thus expect that when the two partitions agree, the algorithm will output the ''correct" partition, while when they disagree, it will compromise between the two, depending on the parameter $\gamma$.

Because of this, we are also able to set $\gamma$ as large as we want (in the interval $[0,1]$, outside of it, it is meaningless), while in the case of the penalty in \eqref{eq:spaOld}, a large value of $\beta$ could result in the entire partition collapsing into a single cluster. This is in particular a problem when the associated network is connected and dense, as we will show in our simulations in Section \ref{sec:sim}. 

We have to be careful with the fact that this method of spatial correction is applicable only to entropic fuzzy clustering and not to the traditional fuzzy clustering where the fuzzyness is tuned by the choice of the exponent $m$ to which memberships are raised in the objective function. Indeed, using modularity correction, it is possible for the argument of the exponential functions in \eqref{eqn:UMd} and \eqref{eqn:UMo} to be negative. This is not  a problem here because the exponential function is always strictly positive, but if the modularity correction was to be applied to non-entropic fuzzy clustering, it could result in negative values of $u_{i,c}$ which are explicitly forbidden.

\section{Validity Measures}
We tested our datasets with a validity measure that takes into account both the network structure and the nodes’ attributes. Specifically, for the fuzzy C-medoids algorithm, similar to \cite{d2024fuzzy}, the index is computed as the sum of the minimum squared distance between medoids and the fuzzy modularity value as defined in \eqref{eq:uCond}, which is in turn divided by an index of within-cluster compactness, as follows:

\begin{equation}\label{eq:valindmedoids}
F(U) = \frac{n-C}{C} \frac{\min\limits_{c\neq c'}d^2(\mathbf{x}_{c},\mathbf{x}_{c'}) + \sum\limits_{n=1}^{N}\sum\limits_{c=1}^{C}\sum\limits_{m=1}^N u_{n,c}b_{n,m}u_{m,c}}{\sum\limits_{n=1}^{N}\sum\limits_{c=1}^{C}u_{n,c}d^2(\mathbf{x}_{n},\mathbf{x}_{c})}
\end{equation}

As for the fuzzy C-modes algorithm, the measure is adjusted so to take into account the Hamming distance between modes (and units), as defined in \eqref{eqn:HamD}. Explicitly, it is computed as:

\begin{equation}\label{eq:valindmodes}
F(U) = \frac{n-C}{C} \frac{\min\limits_{c\neq c'}d_{SM}^2(\mathbf{\hat x}_{c},\mathbf{\hat x}_{c'}) + \sum\limits_{n=1}^{N}\sum\limits_{c=1}^{C}\sum\limits_{m=1}^N u_{n,c}b_{n,m}u_{m,c}}{\sum\limits_{n=1}^{N}\sum\limits_{c=1}^{C}u_{n,c}d_{SM}^2(\mathbf{x}_{n},\mathbf{\hat x}_{c})}
\end{equation}

Since we want the medoids (respectively, the modes) to be well separated and the partition’s fuzzy modularity to be high, we hope for a high value of the numerator; in turn, since the denominator reflects how close the units are to their cluster’s medoid (mode), we hope for a small value. In summary, a higher value of $F$ reflects a better clustering of the data. It’s worth mentioning, though, that by the very structure of our validity measure, a clustering outcome that only takes into account one criterion (network or attributes) but discards the other, will be heavily penalized. It is in principle possible for both validity indices to be negative, however, this requires the fuzzy mdoularity of the partition $\mathbf{U}$ to be negative, that is, the assignment of units to behave worse than the average with respect to the network structure. Such a partition is never desirable for our purposes and thus we  consider this to be a valuable property of the indices.

\section{Simulation study}\label{sec:sim}
\subsection{Fuzzy C-Medoids with modularity spatial correction}\label{subsec:meds}
\begin{figure}[htp]
   \centering
  \includegraphics[width=12cm]{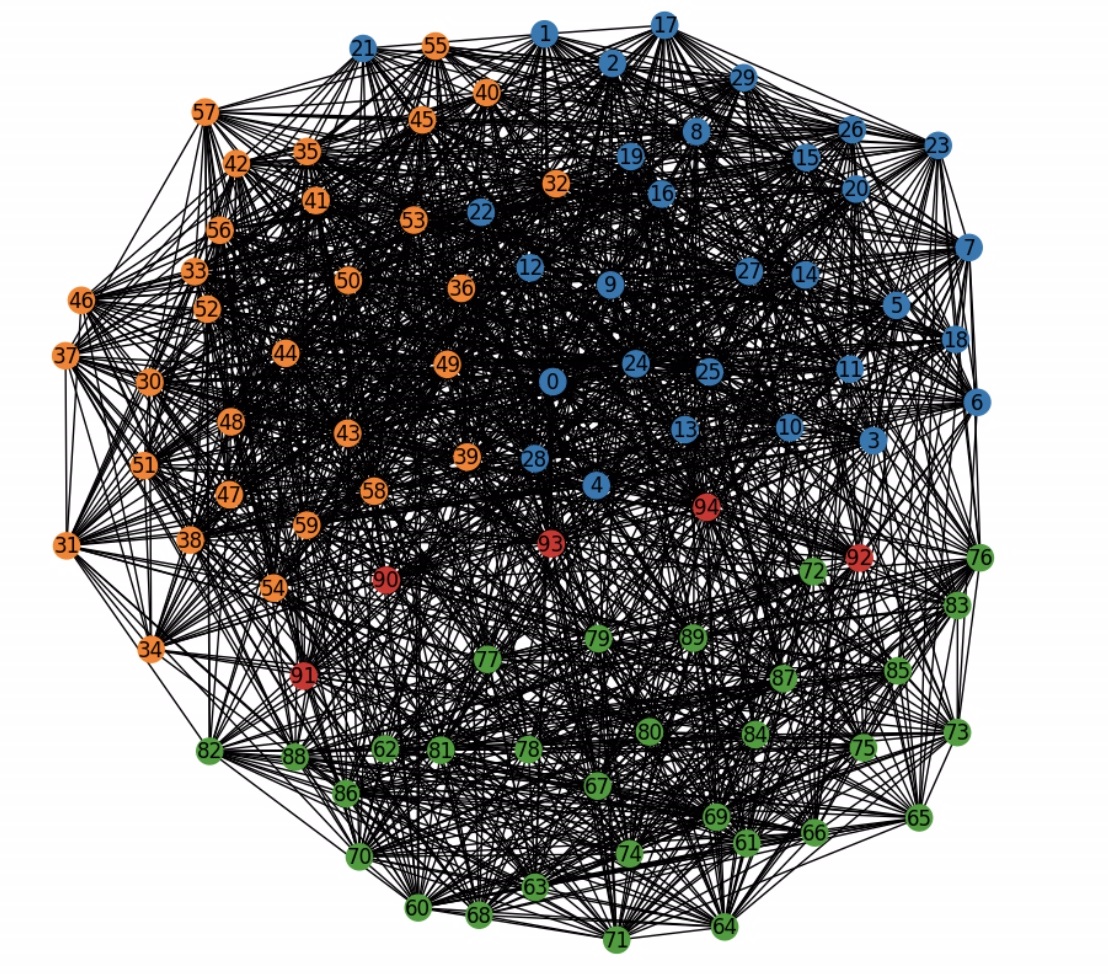}\vspace{0.5cm}
  \includegraphics[width=14cm]{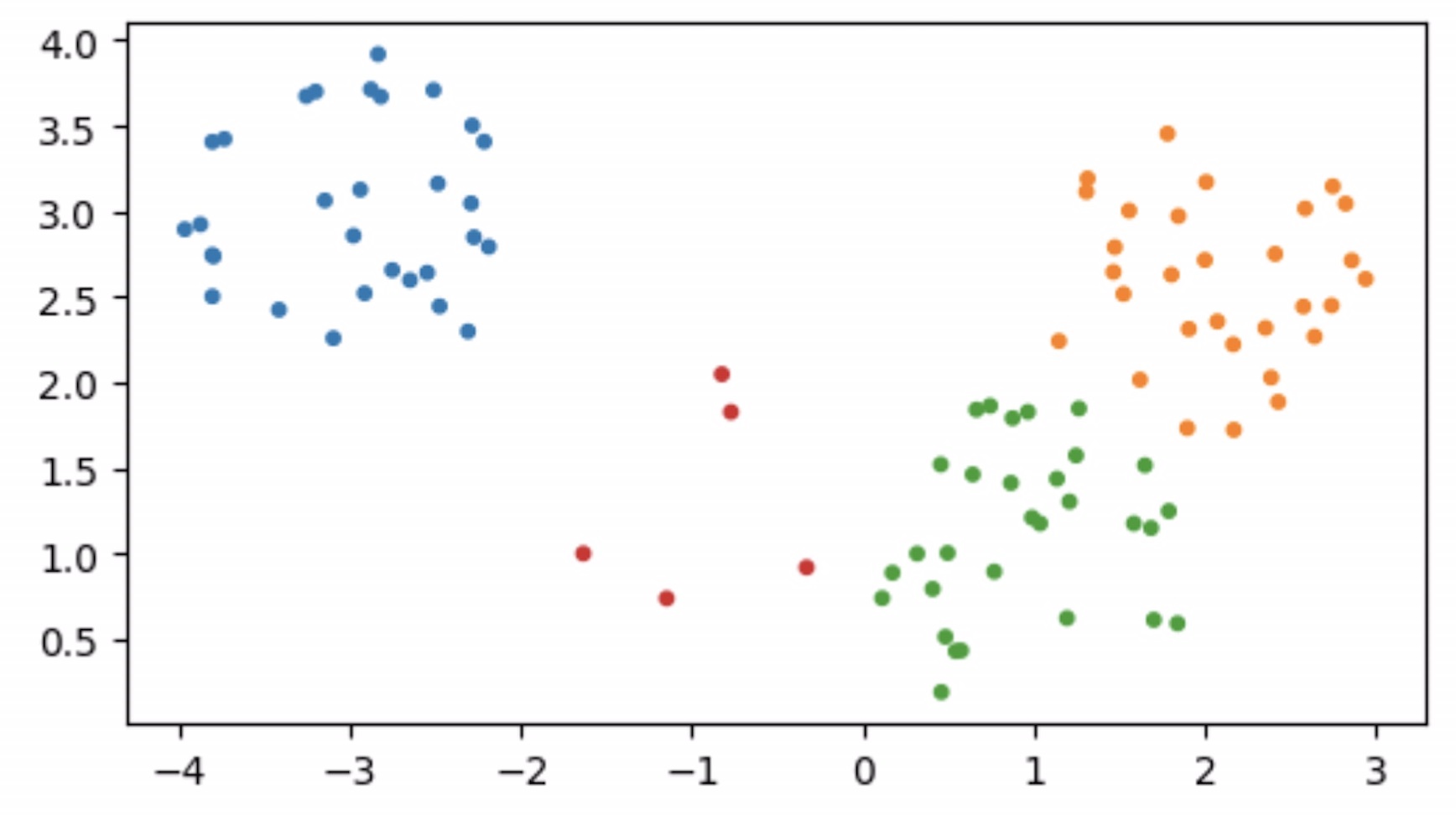}
 \caption{Network structure and attributes for the units in the first scenario. The colors reflect the different clusters.}
 \label{fig:network}
\end{figure}

We carried out a number of simulations in order to test our model's efficiency and hereafter we will discuss the first scenario considered. Specifically, for the FCMd-MSC algorithm, we considered $N=95$ units having $I=2$ attributes grouped in 3 clusters of 30, plus 5 fuzzier ones, in such a way that each cluster was well connected within itself, the first and second clusters were more likely to be connected as per network structure, but more dissimilar as per attributes; in turn, the second and third clusters were closer as far as attributes were concerned, despite being less linked in the network setting. The final 5 units were equally likely to be linked with any other cluster in the network, and set to have attributes close to the average of the remaining ones, as shown in Fig.\ref{fig:network}. 
The graph was generated as a Stochastic Block Model using the \texttt{networkx} python library with probability matrix 

\begin{equation} P = 
\begin{pmatrix}
0.7 & 0.35 & 0.15 & 0.35 \\
0.35 & 0.7 & 0.15 & 0.35 \\
0.15 & 0.15 & 0.7 & 0.35 \\
0.35 & 0.35 & 0.35 & 0.7 .\\
\end{pmatrix}
\end{equation}
Here, the last row and column represent the connection probabilities of the $5$ fuzzy units, while the rest of the matrix describes the connections among the $3$ clusters.
The units' attributes were drawn uniformly from four circles of radius 1 and centers, respectively, $(-3, 3), (2, 2.5), (1, 1)$ and $(-1, 1.5)$. We set the distance $d$ to be the standard Euclidean one in $\mathbb{R}^2$.

\begin{figure}[htp]
    \centering
    \includegraphics[width=12cm]{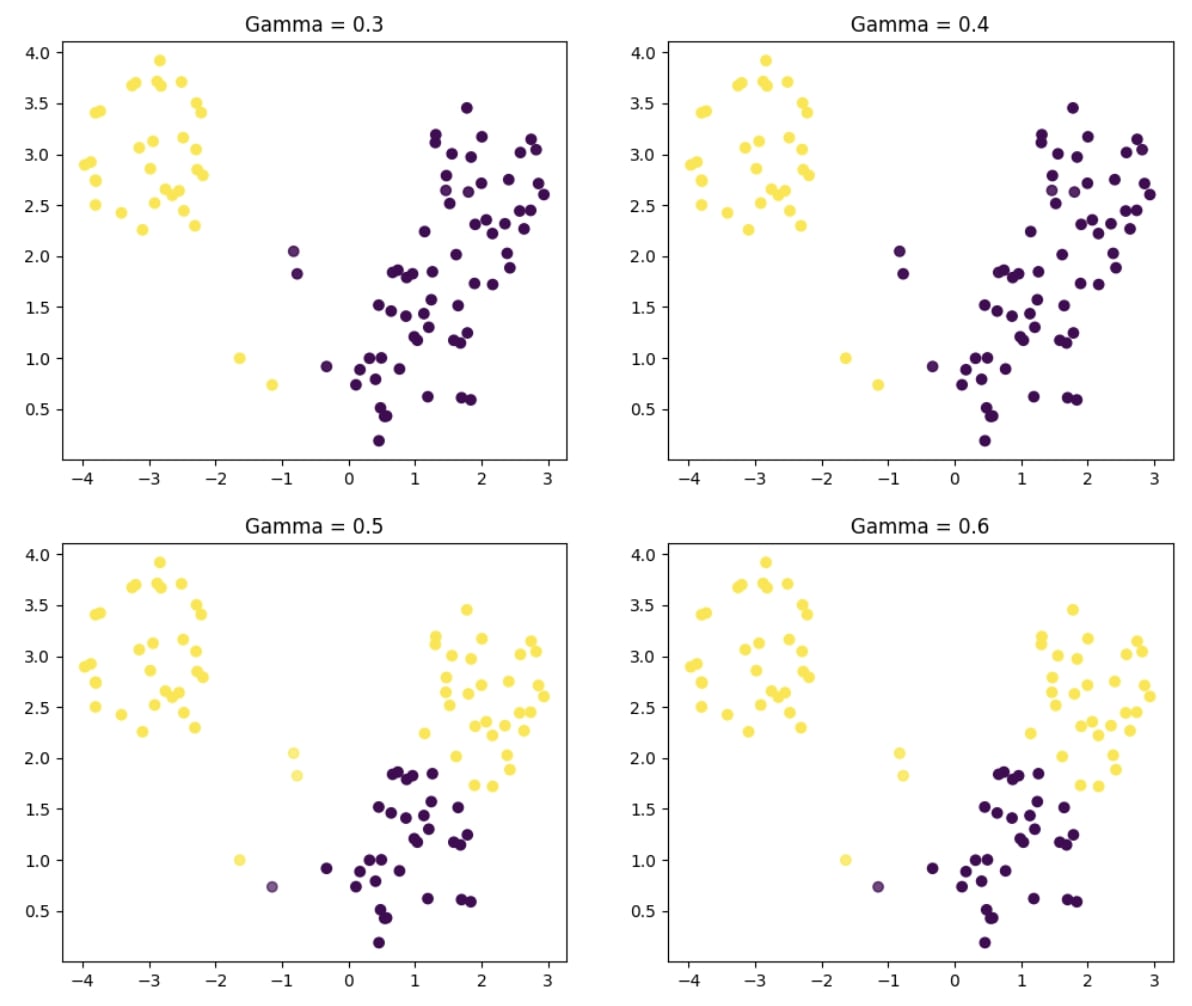}
    \caption{Fuzzy C-Medoids clustering algorithm applied to the dataset described in section \ref{subsec:meds} as $\gamma$ varies, with fixed $C=2$ and $p=0.5$}
    \label{fig:fcmedoids}
\end{figure}

We ran the algorithm with $C=2$, $p=0.5$ and the parameter $\gamma$ ranging from $0.3$ to $0.6$ with a step of $0.1$. As can be seen in Fig.\ref{fig:fcmedoids}, after a certain threshold value $\gamma^\ast \in [0.4,0.5]$ the structure of the clusters changes: for $\gamma < \gamma^\ast$ the attributes' similarity of the units is predominant in the optimization algorithm, whereas for $\gamma \geq \gamma^\ast$ the network structure gains greater importance on the overall clustering criterion. The intensity of the colors, which reflect the units' cluster, is also scaled according to their membership value, thus the fuzzier units appear to be paler than the others.\par
\begin{figure}[htp]
    \centering
    \includegraphics[width=12cm]{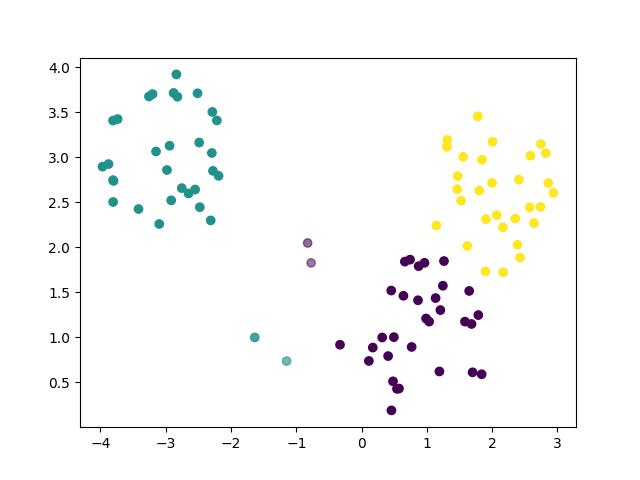}
    \caption{Optimal solution ($C=3$, $\gamma=0.25$) to the Fuzzy C-Medoids clustering algorithm applied to our dataset, with $p=0.5$.}
    \label{fig:optmedoids}
\end{figure}

 
The optimal number of clusters for our dataset was evaluated by means of the validity measure defined in \eqref{eq:valindmedoids}. The values for the validity measure $F(U)$ as the parameters $C$ and $\gamma$ varied can be seen in Table \ref{table:1}, where it can be seen that the optimal combination is $(C, \gamma) = (3, 0.25)$. The corresponding clustering output is shown in Fig.\ref{fig:optmedoids}.

\begin{table}[!ht]
\centering
\resizebox{\columnwidth}{!}{%
\begin{tabular}{l|rrrrrrrrrrrrr}
\toprule
&&&&&&$\gamma$&&&&&&& \\
 $C$& 0 & 0.05 & 0.10 & 0.15 &0.20 &0.25 &0.30 &0.35 &	0.40 &	0.45 &	0.50 &	0.55 &	0.60 \\
\midrule
2 & 211.0 & 213.5 & 215.1 & 216.2 & 216.9 & 217.3 & 217.4 & 217.1 & 216.3 & 214.6 & 167.4 & 167.5 & 167.5 \\
3 & 237.4 & 308.9 & 336.3 & 342.2 & 343.9 & \textbf{344.3} & 344.3 & 344.0 & 343.8 & 343.6 & 343.4 & 343.2 & 343.1 \\
4 & 131.7 & 164.3 & 222.6 & 271.3 & 275.0 & 276.4 & 276.7 & 276.1 & 275.5 & 274.8 & 274.0 & 273.1 & 269.2 \\
5 & 76.7 & 104.5 & 132.3 & 212.8 & 211.5 & 212.5 & 216.6 & 219.4 & 219.2 & 216.4 & 218.7 & 212.9 & 218.0 \\
\end{tabular}%
}
\caption{Validity measures for FCMd-MSC model corresponding to varying $C$ and $\gamma$ values and fixed $p=0.5$.}
\label{table:1}
\end{table}

To further highlight the benefits of our proposed method, we tested the same dataset described in Section \ref{subsec:meds} with a different Fuzzy C-Medoids algorithm in which the spatial term defined in \eqref{eq:spaOld} penalizes a partition grouping adjacent units in different clusters. In particular, the model aimed at minimizing the function 

\begin{align}\label{eq:CMedOld}\nonumber
J(\mathbf{U}, \mathbf{x}_c) =(1-\beta)\sum\limits_{n=1}^{N}&\sum\limits_{c=1}^{C}u_{n,c}d^2(\mathbf{x}_{n},\mathbf{x}_{c})+p\sum_{n=1}^N\sum\limits_{c=1}^{C}u_{n,c}\log (u_{n,c})\\&+\frac{\beta}{2}\sum\limits_{n=1}^{N}\sum\limits_{c=1}^{C}u_{n,c} \sum\limits_{m=1}^{N}\sum\limits_{{c'\in C_c}}a_{n,m}u_{m,c'}
\end{align}

\begin{figure}[htp]
    \centering
    \includegraphics[width=12cm]{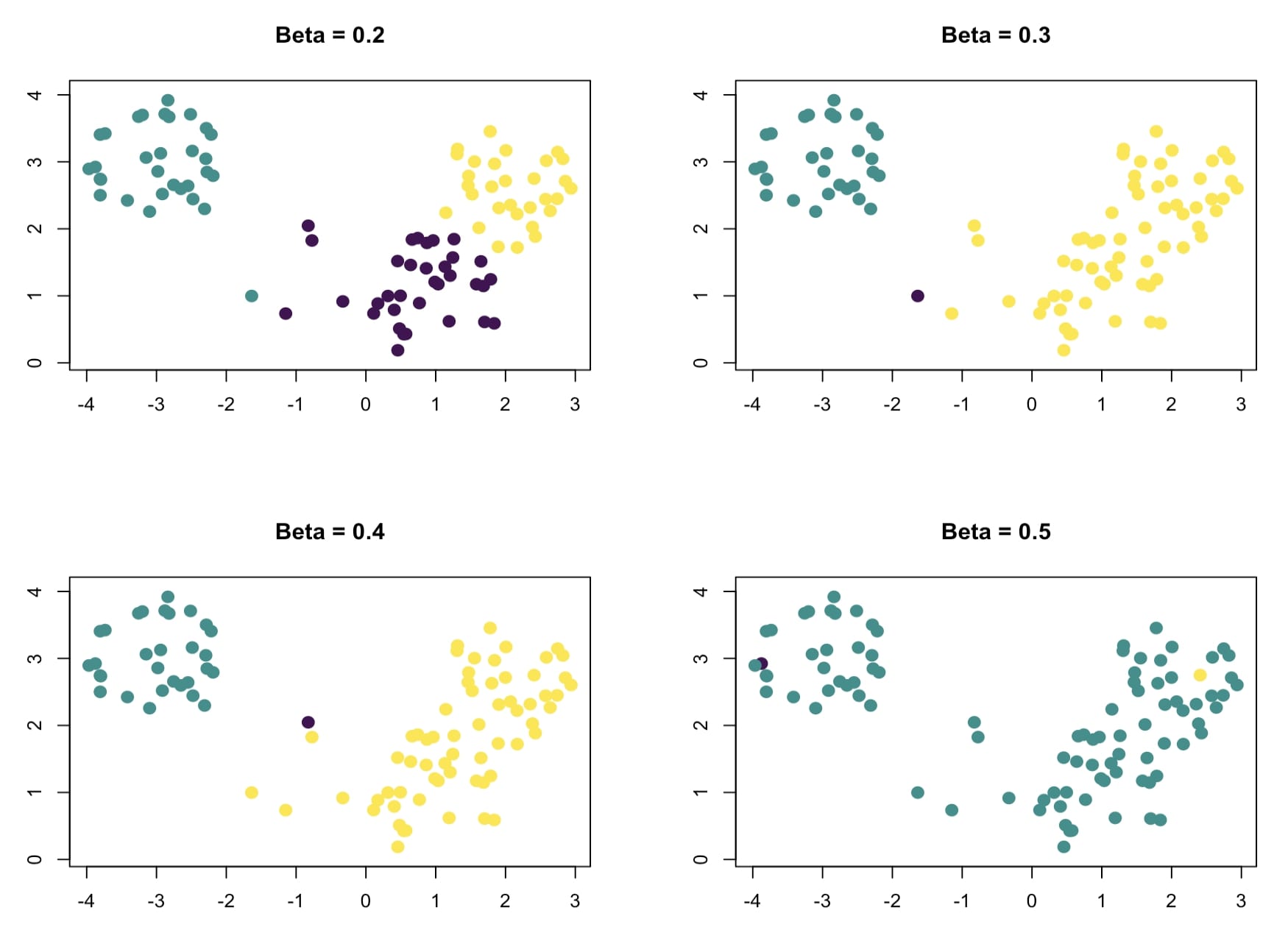}
    \caption{Fuzzy C-Medoids model with spatial penalty tuned by the parameter $\beta$ applied to our dataset, with $p=0.5$, $C=3$.}
    \label{fig:oldmedoids}
\end{figure}

As already discussed in Section \ref{subsec:rationale}, since this model does not, in turn, penalize a partition grouping non-adjacent units into the same cluster, increasing the tuning parameter $\beta$ will eventually result in all the clusters collapsing into a single one, as shown in Fig.\ref{fig:oldmedoids}.

\subsection{Fuzzy C-Modes with modularity spatial correction}\label{subsec:modes}
\begin{figure}[htp]
    \centering
    \includegraphics[width=8cm]{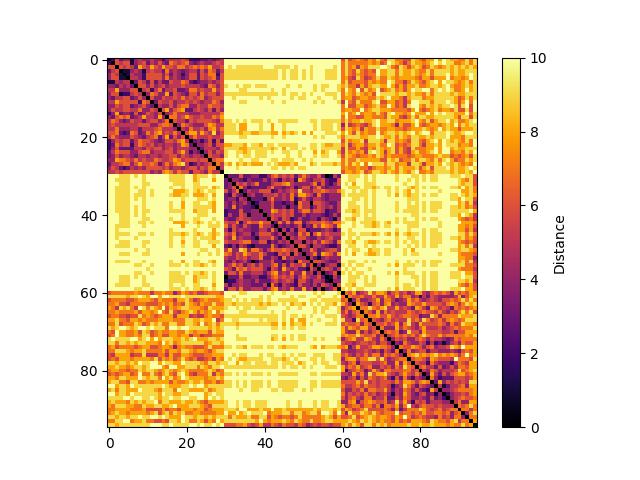}
    \caption{Hamming distance of the units for FCMo-MSC algorithm testing.}
    \label{fig:distmatrix}
\end{figure}

In order to test the FCMo-MSC algorithm's efficiency, we created a dataset following the same idea as for its C-Medoids counterpart. We considered $N=95$ units having $I=10$ attributes. The network structure was the same as the one described in section \ref{subsec:meds} (and shown in Fig.\ref{fig:network}), whereas the attributes were created sampling from the set $\Omega = \{ A,B,C,D,E \}$ according to the following criteria:
\begin{itemize}
    \item $n=1,...,30$: attributes $x_{n,i}$ were drawn from the set $\Omega$ with probability vector given by the $i_{\mathrm{th}}$ column of the following table:
    
    \begin{table}[!ht]
    \centering
    \resizebox{0.8\columnwidth}{!}{%
    \begin{tabular}{l|rrrrrrrrrr}
    \toprule
     & $x_{n,1}$ & $x_{n,2}$ & $x_{n,3}$ & $x_{n,4}$ &$x_{n,5}$ &$x_{n,6}$ & $x_{n,7}$ & $x_{n,8}$ & $x_{n,9}$ &$x_{n,10}$ \\
    \midrule
    $A$&0.6 &0.6& 0.1 &0.1 &0 & 0 &0.1 &0.1 &0 & 0 \\
    $B$&0.3 &0.3 &0.6 &0.6 &0 & 0 & 0  &0 & 0.3& 0.3\\
    $C$&0.1 &0.1& 0.2& 0.2& 0.1& 0.1& 0.6& 0.6& 0.6& 0.6\\
    $D$&0 & 0&  0.1& 0.1 &0.6& 0.6& 0.1& 0.1& 0.1& 0.1\\
    $E$&0 & 0&  0&  0&  0.3& 0.3& 0.2 &0.2& 0 & 0\\
    \end{tabular}%
    }
    \end{table}
    
    \item $n=31,...,60$: attributes $x_{n,i}$ were drawn from the set $\Omega$ with probability vector given by the $i_{\mathrm{th}}$ column of the following table:

    \begin{table}[!ht]
    \centering
    \resizebox{0.8\columnwidth}{!}{%
    \begin{tabular}{l|rrrrrrrrrr}
    \toprule
     & $x_{n,1}$ & $x_{n,2}$ & $x_{n,3}$ & $x_{n,4}$ &$x_{n,5}$ &$x_{n,6}$ & $x_{n,7}$ & $x_{n,8}$ & $x_{n,9}$ &$x_{n,10}$ \\
    \midrule
    $A$ &0   &0  &0   &0  &0.7 &0.7 &0.3 &0.3 &0.7 &0.7 \\
    $B$ &0   &0  &0   &0  &0   &0   &0.7 &0.7 &0   &0   \\
    $C$ &0.3 &0.3&0   &0  &0.3 &0.3 &0   &0   &0   &0   \\
    $D$ &0   &0  &0.7 &0.7&0   &0   &0   &0   &0.3 &0.3 \\
    $E$ &0.7 &0.7&0.3 &0.3&0   &0   &0   &0   &0   &0   \\
    \end{tabular}%
    }
    \end{table}
    
    \item $n=61,...,90$: attributes $x_{n,i}$ were drawn from the set $\Omega$ with probability vector given by the $i_{\mathrm{th}}$ column of the following table:
    
    \begin{table}[!ht]
    \centering
    \resizebox{0.8\columnwidth}{!}{%
    \begin{tabular}{l|rrrrrrrrrr}
    \toprule
     & $x_{n,1}$ & $x_{n,2}$ & $x_{n,3}$ & $x_{n,4}$ &$x_{n,5}$ &$x_{n,6}$ & $x_{n,7}$ & $x_{n,8}$ & $x_{n,9}$ &$x_{n,10}$ \\
    \midrule
    $A$ &0.3 &0.3 &0.6  &0.6  &0   &0   &0.1 &0.1 &0   &0   \\
    $B$ &0.1 &0.1 &0.2  &0.2  &0.6 &0.6 &0   &0   &0   &0   \\
    $C$ &0   &0   &0.1  &0.1  &0.1 &0.1 &0.2 &0.2 &0.3 &0.3 \\
    $D$ &0.6 &0.6 &0.1  &0.1  &0.3 &0.3 &0.6 &0.6 &0.1 &0.1 \\
    $E$ &0   &0   &0    &0    &0   &0   &0.1 &0.1 &0.6 &0.6 \\
    \end{tabular}%
    }
    \end{table}

    \item $n=91,...,95$: attributes $x_{n,i}$ were drawn from the set $\Omega$ with probability vector given by the $i_{\mathrm{th}}$ column of the following table:

    \begin{table}[!ht]
    \centering
    \resizebox{0.8\columnwidth}{!}{%
    \begin{tabular}{l|rrrrrrrrrr}
    \toprule
     & $x_{n,1}$ & $x_{n,2}$ & $x_{n,3}$ & $x_{n,4}$ &$x_{n,5}$ &$x_{n,6}$ & $x_{n,7}$ & $x_{n,8}$ & $x_{n,9}$ &$x_{n,10}$ \\
    \midrule
    $A$ &0.3 &0.3 &0.3  &0.3  &0.3 &0.3 &0.1 &0.1 &0.3 &0.3 \\
    $B$ &0.1 &0.1 &0.3  &0.3  &0.3 &0.3 &0.3 &0.3 &0.1 &0.1 \\
    $C$ &0   &0   &0.1  &0.1  &0   &0   &0.3 &0.3 &0.3 &0.3 \\
    $D$ &0.3 &0.3 &0.3  &0.3  &0.3 &0.3 &0.3 &0.3 &0   &0   \\
    $E$ &0.3 &0.3 &0    &0    &0.1 &0.1 &0   &0   &0.3 &0.3 \\
    \end{tabular}%
    }
    \end{table}
\end{itemize}
The distance matrix of the units, computed according to \eqref{eqn:HamD}, is shown in Fig.\ref{fig:distmatrix} and reflects the closeness of the first and third blocks; we recall that, as per network structure, the first and second blocks were set to be more likely to be linked to each other than to the third one. 

\begin{figure}[htp]
    \centering
    \includegraphics[width=12cm]{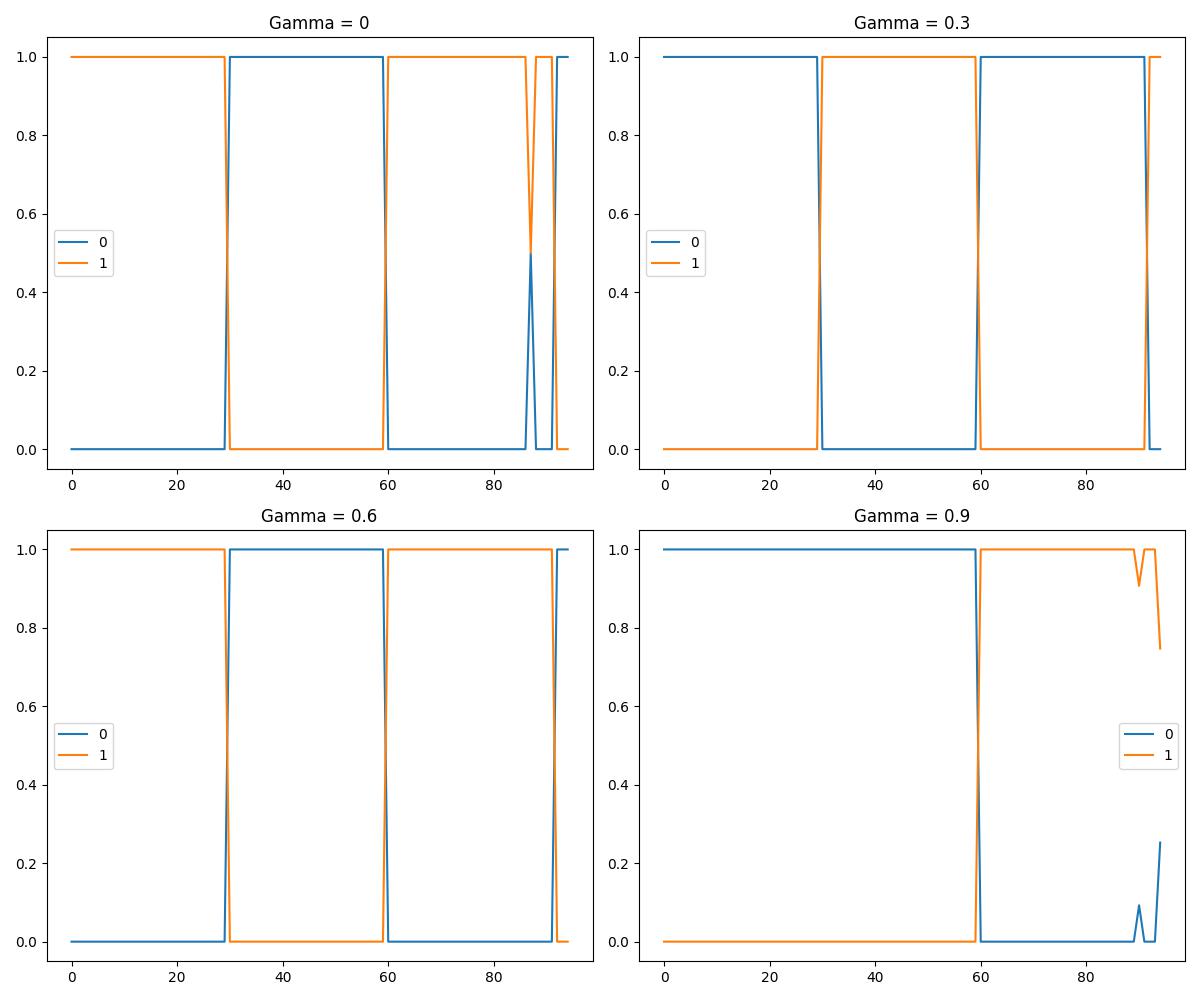}
    \caption{Plot of the $\mathbf{U}$ matrices for increasing values of $\gamma$, $C=2$ and $p=0.2$. The $x$ axis indicates the unit label and each line the membership value to a cluster.}
    \label{fig:fcmodes}
\end{figure}

We ran the algorithm with $C=2$ and increasing values of $\gamma$, thus incrementing the network's importance in the clustering criterion. Fig.\ref{fig:fcmodes} reports the membership values of the units (i.e. the matrix $\mathbf{U}$) to their cluster. As in the FCMd-MSC context, we can infer the existence of a threshold value $\gamma^\ast$ before which the units' similarities prevail over their adjacencies (so that units from the first and third block get high membership values to the same cluster), and after which the opposite occurs (so that units from the first and second block get high membership values to the same cluster).

\begin{figure}[htp]
    \centering
    \includegraphics[width=12cm]{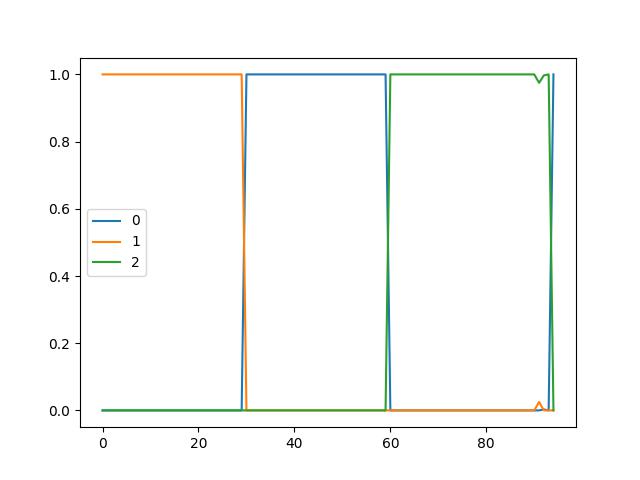}
    \caption{Membership plot for the optimal solution ($C=3$, $\gamma=0.3$) to the Fuzzy C-Modes clustering algorithm applied to our dataset, with $p=0.2$. The $x$ axis indicates the unit label and each line the membership value to a cluster.}
    \label{fig:optmodes}
\end{figure}

We tested the clustering efficiency of our model by running the algorithm with varying values of $C$ and $\gamma$ and computing, for each combination, the validity measure defined in \eqref{eq:valindmodes}. As shown in Table \ref{table:2}, the validity index reaches its maximum for $(C, \gamma) = (3, 0.3)$, as we expected by the dataset structure. The membership plot corresponding to the optimal combination is shown in Fig.\ref{fig:optmodes}. 

\begin{table}[!ht]
\centering
\resizebox{\textwidth}{!}{%
\begin{tabular}{l|rrrrrrrrrrrrr}
\toprule
&&&&&&$\gamma$&&&&&&& \\
 $C$& 0 & 0.05 & 0.10 & 0.15 &0.20 &0.25 &0.30 &0.35 &	0.40 &	0.45 &	0.50 &	0.55 &	0.60 \\
\midrule
2 & 8.20 & 8.42 & 8.44 & 8.44 & 8.44 & 8.44 & 8.44 & 8.44 & 8.44 & 8.44 & 8.44 & 8.44 & 8.44 \\
3 & 16.13 & 16.32 & 16.39 & 16.42 & 16.43 & 16.43 & \textbf{16.44} & 16.44 & 16.44 & 16.44 & 16.44 & 16.44 & 16.44 \\
4 & 10.11 & 10.87 & 11.48 & 10.51 & 11.55 & 11.12 & 10.93 & 10.93 & 11.58 & 11.28 & 12.60 & 11.68 & 12.77 \\
5 & 7.90 & 7.52 & 8.79 & 7.89 & 8.42 & 8.51 & 9.29 & 7.96 & 7.56 & 8.72 & 9.57 & 9.14 & 8.85 \\
\end{tabular}%
}
\caption{Validity measures for FCMo-MSC model corresponding to varying $C$ and $\gamma$ values and fixed $p=0.2$.}
\label{table:2}
\end{table}

The same dataset was tested using another Fuzzy C-Modes clustering algorithm in order to assess its performance when compared to our proposed model. The model was based on finding the optimal partition and modes that minimized the function 

\begin{align}\label{eq:CModOld}\nonumber
F(\mathbf{U}, \mathbf{\hat x}_c) = \sum\limits_{n=1}^{N}&\sum\limits_{c=1}^{C}u_{n,c} d_{SM}^2(\mathbf{x}_{n},\mathbf{\hat x}_{c}) +p\sum_{n=1}^N\sum\limits_{c=1}^{C}u_{n,c}\log (u_{n,c})\\&+\frac{\beta}{2}\sum\limits_{n=1}^{N}\sum\limits_{c=1}^{C}u_{n,c}
	\sum\limits_{m=1}^{N}\sum\limits_{{c'\in C_c}}a_{n,m}u_{m,c'}
\end{align}

\begin{figure}[htp]
    \centering
    \includegraphics[width=12cm]{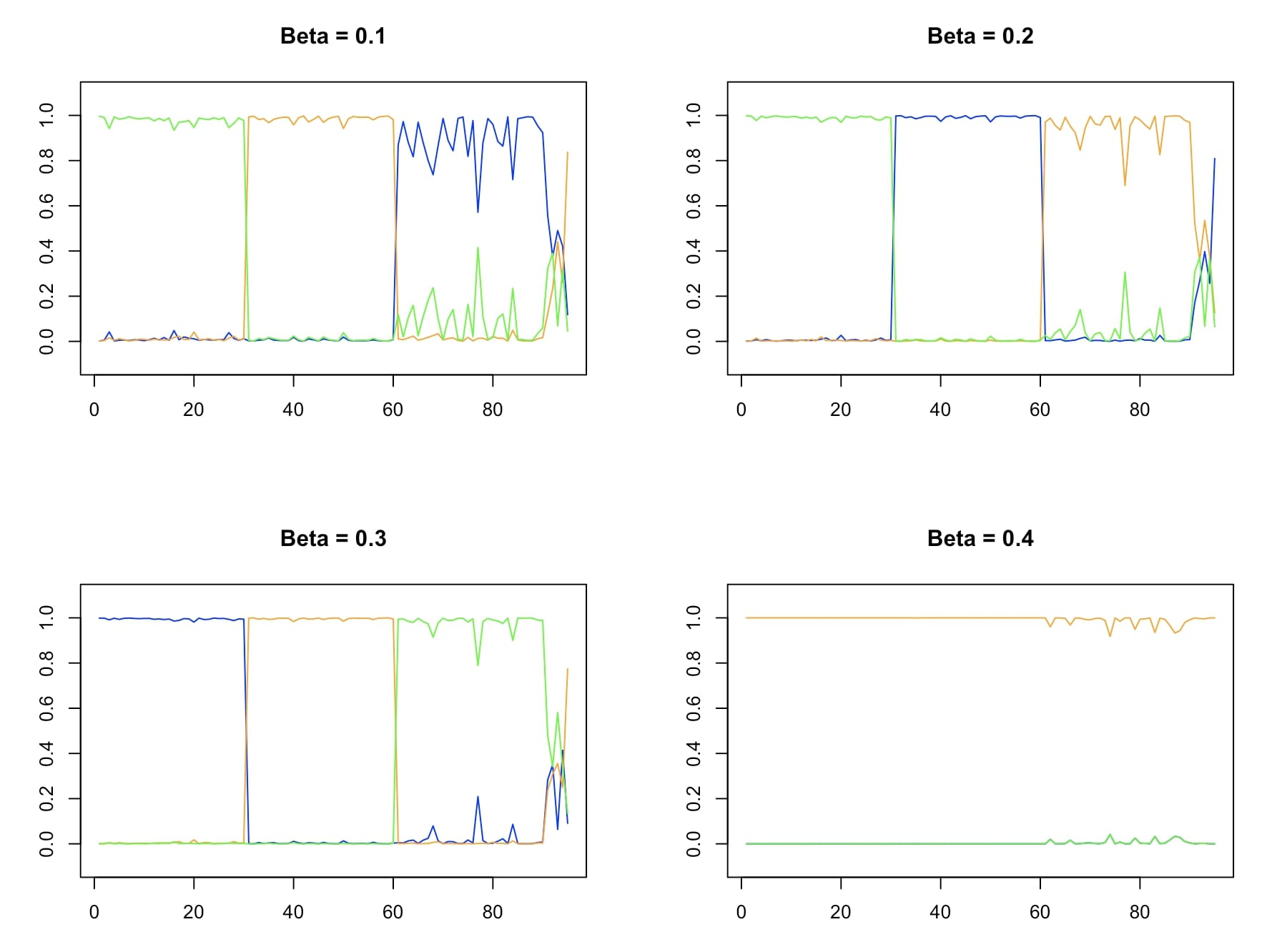}
    \caption{The units' membership values when clustered according to \eqref{eq:CModOld}, with $p=1.5$ and $C=3$.}
    \label{fig:oldmodes}
\end{figure}

Again, since the spatial term does not penalize a partition in which one cluster contains non-adjacent units, when increasing the weight $\beta$ all the clusters will eventually collapse into a single one. This is shown in Fig.\ref{fig:oldmodes}, where we can observe that after a certain threshold value of $\beta$, all the units have highest membership value to the same cluster.

\section{Application of FCMd-MSC to 2030 Agenda for Sustainable Development}
\label{sec:AppMd}
We tested the FCMd-MSC algorithm on real data, focusing on the indicators proposed by \cite{sachs2023} and related to the 16th goal of the 2030 Agenda for Sustainable Development.  This goal focuses on promoting peace, reducing violence, ensuring equal access to justice for all, building effective and accountable institutions, and promoting inclusive societies. It emphasizes the importance of peace and stability as prerequisites for sustainable development.
The indicators used by \cite{sachs2023} are listed below.
\begin{description} 
\item[Homicides:] Homicides (per 100,000 population)
\item[Detain:] Unsentenced detainees (\% of prison population)
\item[Safe:] Population who feel safe walking alone at night in the city or area where they live (\%)
\item[U5 Registration:] Birth registrations with civil authority (\% of children under age 5)
\item[CPI:] Corruption Perceptions Index 
\item[Child Labor:] Children involved in child labor (\% of population aged 5 to 14)
\item[Weapons Expenditure:] Exports of major conventional weapons (TIV constant million USD per 100,000 population)
\item[RSF:] Press Freedom Index 
\item[Justice:] Access to and affordability of justice 
\item[Admin:] Timeliness of administrative proceedings
\item[Expropriation risk:] Expropriations are lawful and adequately compensated 
\end{description} 
As outlined by \cite{sachs2023}, the indicators were normalized on a scale from 0 (worst performance) to 100 (best performance). Only nations with complete values for these indicators were included in the analysis, resulting in a dataset of 78 countries. 

The adjacency matrix was derived from data in the 2024 passport-index-dataset repository \href{https://github.com/ilyankou/passport-index-dataset/blob/master/README.md}{(passport-index-dataset)}, where a value of 1 indicates that travel is possible between two countries without a visa (at least temporarily). 

We conducted the analysis with $p = 3$, $C \in \lbrace 2,3,4,5 \rbrace$, and $\gamma \in [0,1]$ with an increment of $0.1$. The number of random restarts was set to 50, with a maximum of 3000 iterations. Based on the proposed validity index, the optimal solution was found at \( C = 2 \) and \( \gamma = 0.7 \), as shown in Table \ref{tab:gamma}.
\begin{table}[!ht]
    \centering\resizebox{\columnwidth}{!}{
\begin{tabular}{l|rrrrrrrrrrr}
\toprule
&&&&&&$\gamma$&&&&& \\
C & 0 & 0.1 & 0.2 & 0.3 & 0.4 & 0.5 & 0.6 & 0.7 & 0.8 & 0.9 & 1 \\
\midrule
2&	2.85&	2.89&	2.94&	3.00&	3.18&	3.28&	3.35&	\textbf{3.36}&	3.07&	2.52&	2.25\\
3&	1.77&	1.79&	1.47&	1.49&	1.51&	1.52&	1.51&	1.48&	1.36&	0.71&	0.61\\
4&	1.30&	1.31&	1.19&	1.20&	1.21&	0.77&	0.74&	0.74&	0.44&	0.24&	0.12\\
5&	0.84&	0.81&	0.81&	0.80&	0.69&	0.71&	0.47&	0.39&	0.23&	0.12&	0.08\\
\bottomrule
\end{tabular}}
    \caption{Validity measures for FCMd-MSC with $p = 3$, $C \in \lbrace 2,3,4,5 \rbrace$, and $\gamma \in [0,1]$ with step $0.1$}
    \label{tab:gamma}
\end{table}
The resulting partition shows that the first cluster, with Guyana as its medoid, includes 39 countries, while the second cluster, with  Latvia as its medoid, consists of 33 countries. Six countries are classified as fuzzy units, since they do not show a membership degree $u_{ic}\geq 0.7$ to either cluster.

\begin{figure}
\includegraphics[width=0.9 \textwidth]{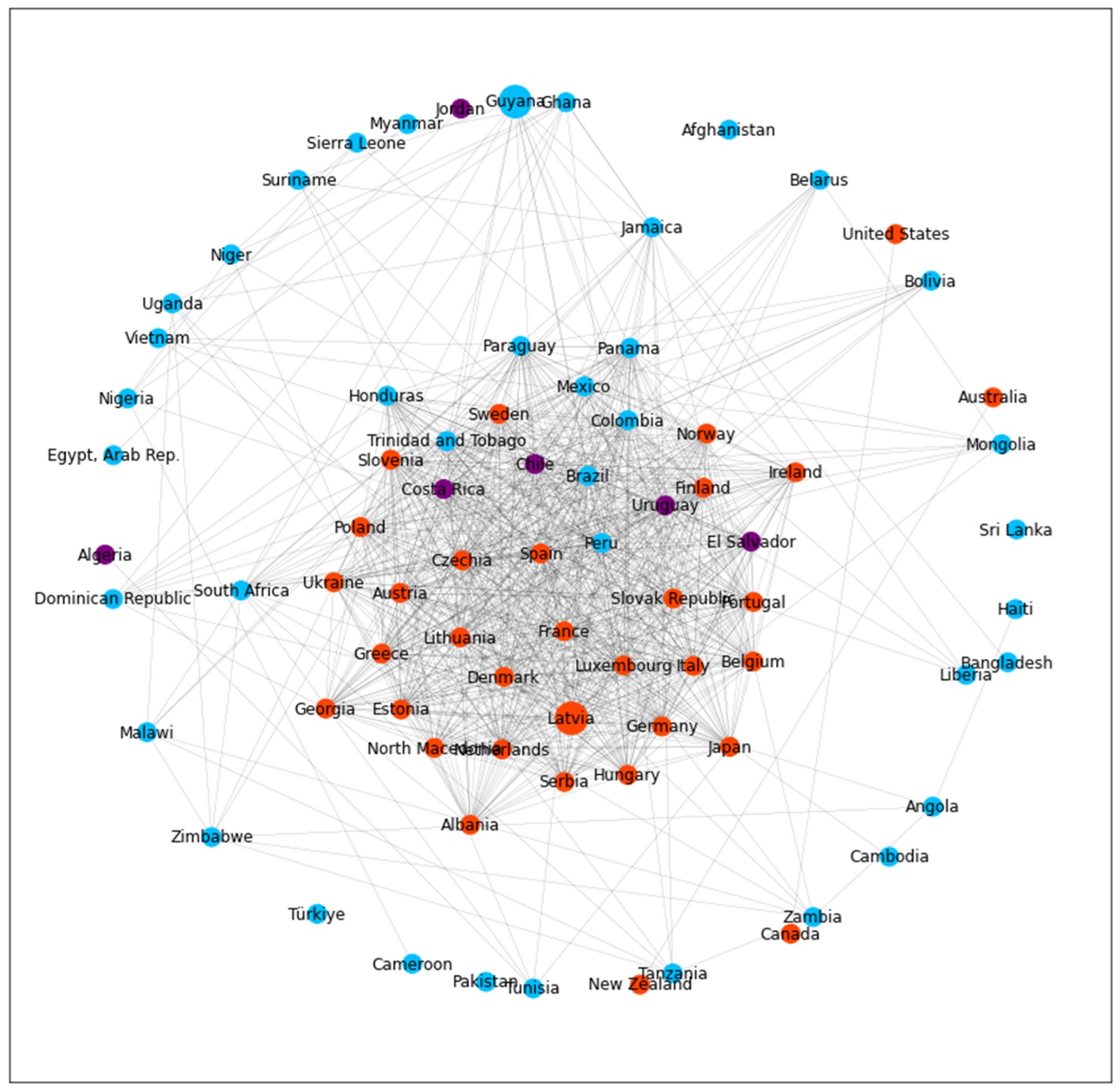}
\caption{Network based on the 2024 passport-index-dataset repository, with nodes colored by cluster membership. Fuzzy nodes are highlighted in purple. Bigger dots represent medoids.}
    \label{network}
\end{figure}

The obtained clusters can be visualized in Fig \ref{network} which displays nodes colored by cluster membership; nodes shown in cyan represent fuzzy units. The membership degrees matrix $U$ is provided in Table \ref{partitions0} in Appendix \ref{app:longtables}.

\begin{table}[ht]
\centering
\begin{tabular}{l|c|c|c}
\hline
\textbf{Indicator} & \textbf{Cluster 1} & \textbf{Cluster 2} & \textbf{Fuzzy units} \\
\hline
Homicides & 71.79 & 97.08 & 81.03 \\
Detain & 50.09 & 74.64 & 72.30 \\
Safe & 34.50 & 74.91 & 42.69 \\
U5 Registration & 72.30 & 99.88 & 97.68 \\
CPI & 25.91 & 68.49 & 50.71 \\
Child Labor & 64.57 & 98.42 & 89.41 \\
Weapons Expenditure & 98.85 & 78.59 & 97.75 \\
RSF & 28.43 & 76.35 & 38.37 \\
Justice & 59.95 & 85.16 & 82.67 \\
Admin & 33.27 & 70.83 & 53.31 \\
Expropriation Risk & 18.81 & 64.75 & 50.97 \\
\hline
\end{tabular}
\caption{Average Indicator Scores by Cluster}
\label{average}
\end{table}

As shown in Table \ref{average}, which reports the mean values of indicators by cluster, countries in Cluster 1 exhibit higher levels of insecurity, lower institutional effectiveness, and limited press freedom. In contrast, countries in Cluster 2 generally perform better across all indicators (except for Weapons Expenditure, which is typically higher in wealthier countries), displaying stronger institutional scores and lower levels of violence and corruption. Fuzzy units show mixed performance on these indicators. Although they perform better than Cluster 1, there are some areas for improvement.

Considering the visa requirement and hence looking at the network structure, we observe that countries with higher scores on the indicators, \textit{i.e.} mainly those in Cluster 2, generally belong to areas where travel is possible without a visa, especially within the European Union. This facilitates cooperation and movement for work and study purposes, further strengthening security and economic stability.

On the other hand, in countries with generally lower scores on the indicators, \textit{i.e.} mainly those in Cluster 1, visa requirements or restrictions are more common, as expected in regions with greater political instability.
 These barriers limit mobility and, indirectly, influence trade and cultural exchange.

\section{Application of FCMo-MSC to urban green spaces in Italian provincial capital and metropolitan cities}
\label{sec:appMo}
We tested the FCMo-MSC algorithm on real data concerning urban green spaces in the provincial and metropolitan capitals of Italy provided by Istat as part of the "Urban Environment" survey for 2022. This annual survey collects and analyzes data on the urban environment in Italian cities, including the 109 provincial or metropolitan capitals and the Municipality of Cesena, which participates voluntarily. Since 2000, the survey has focused on tracking various environmental and service-related aspects within urban areas, thereby supporting evaluations of quality of life and local environmental policies.
The survey encompasses eight key topics: water, air, environmental management, energy, urban waste, noise, urban mobility and green spaces.
Focusing on the last topic, for each city, we considered the following $9$ qualitative attributes, some of which were derived by aggregating original variables:

\begin{description}
    \item[A1] "Urban Green Census" with categories: "No","Yes, part of the municipal area","Yes, the entire municipal area"
    \item[A2] "Urban Green Census with Georeferenced Data" with categories: "No","Yes, part of the municipal area","Yes, the entire municipal area"
    \item[A3] "Urban Green Census with the Green Information System" with categories: "No", "Yes"
    \item[A4] "Publication of the Tree Balance Sheet as of December 31, 2022" with categories: "No", "Yes"
    \item[A5] "Monitoring of the Risk of Roadside Tree Failure as of December 31, 2022" with categories: "No", "Yes"
    \item[A6] "Greening of Areas Subject to New Construction or Significant Renovation" with categories "No","With incentives and verification by private entities","Yes, with direct action by the municipality","Both"
    \item[A7] "Increase, Conservation, and Protection of Arboreal Heritage in Open Areas Adjacent to Existing Buildings" with categories "No", "With incentives and verification by private entities","Yes, with direct action by the municipality", "Both"
    \item[A8] "Local Initiatives for the Maintenance and Management of Urban Green Spaces Assigned to Citizens or Associations Free of Charge by Municipal Administrations" with categories: "No", "Yes"
    \item[A9] "Areas designated for urban afforestation" with categories: "No", "Yes"
\end{description}

The spatial relationships among the $110$ provincial and metropolitan city capitals are encoded in the contiguity matrix $\mathbf{A}_{110 \times 110}$, where the element $a_{ii'}$ is equal to 1 if and only if the distance between locations $i$ and $i'$ is less than or equal to 100 km.

We run the algorithm considering $p=1$, $C \in \lbrace 2,3,4,5\rbrace$, $\gamma \in [0,1]$ with step $0.1$. We fixed the number of random restarts to $25$  and the maximum number of iterations to $1000$.
Based on the proposed validity index, the optimal solution is identified for $C = 2$ and $\gamma = 0.9$, as shown in Table \ref{tab:v_i}. However, we also considered the best solution for  $C = 3$, which shares the same $\gamma$ value. This consideration allows us to explore the potential advantages of a more complex clustering configuration while maintaining a consistent evaluation criterion.
\begin{table}[!ht]
    \centering\resizebox{\columnwidth}{!}{
\begin{tabular}{l|rrrrrrrrrrr}
\toprule
&&&&&&$\gamma$&&&&& \\
C & 0 & 0.1 & 0.2 & 0.3 & 0.4 & 0.5 & 0.6 & 0.7 & 0.8 & 0.9 & 1 \\
\midrule
2 & 5.42 & 6.05 & 6.67 & 5.91 & 6.12 & 6.53 & 7.59 & 9.59 & 13.30 & \textbf{13.72} & 11.57 \\
3 & 2.46 & 3.96 & 4.30 & 4.64 & 5.04 & 5.67 & 8.03 & 10.61 & 13.03 & 13.18 & 12.37 \\
4 & 1.82 & 1.97 & 2.73 & 3.73 & 3.52 & 5.30 & 5.20 & 8.44 & 9.67 & 10.19 & 8.38 \\
5 & 1.34 & 2.38 & 2.44 & 2.25 & 1.94 & 3.21 & 4.38 & 7.11 & 8.30 & 8.43 & 7.32 \\
\bottomrule
\end{tabular}}
    \caption{Validity measures for FCMo-MSC model corresponding to varying $C$ and $\gamma$ values and fixed $p=1$.}
    \label{tab:v_i}
\end{table}

\begin{figure}[H]
    \centering
    \subfloat[]{
        \includegraphics[scale=0.6]{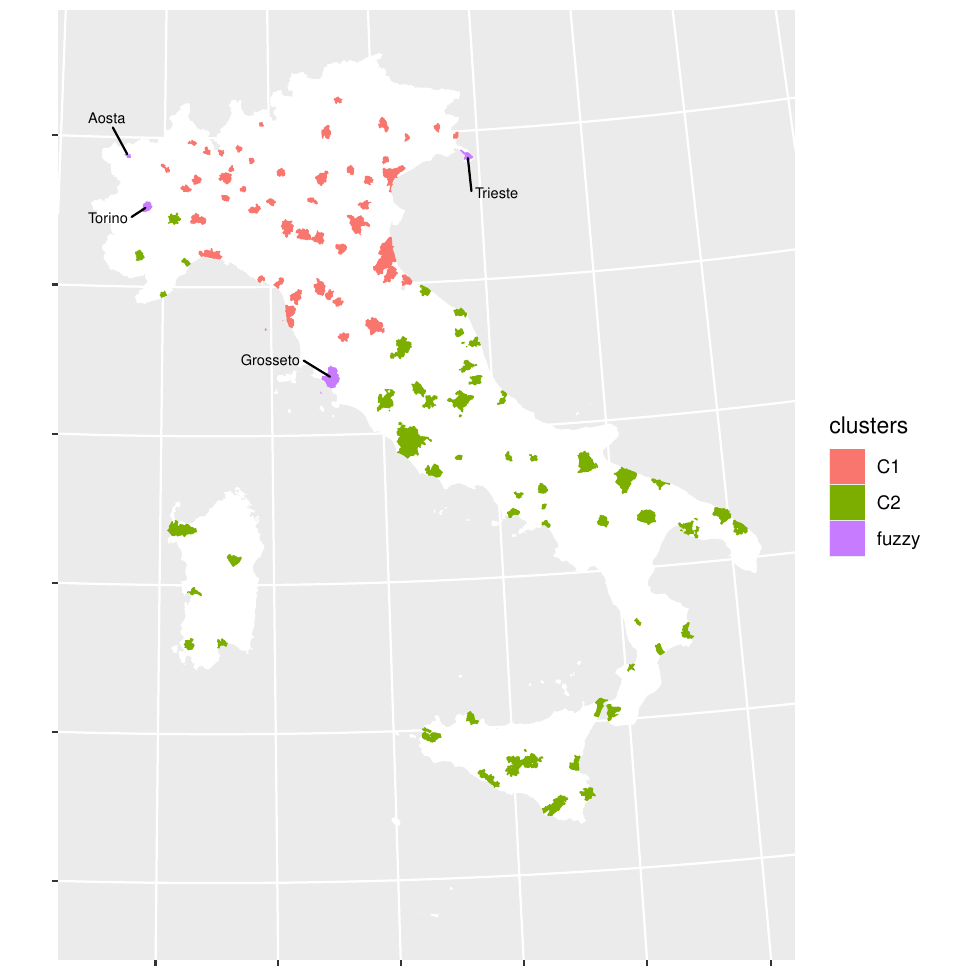}
    } \vspace{0.5cm}
    \subfloat[]{
        \includegraphics[scale=0.6]{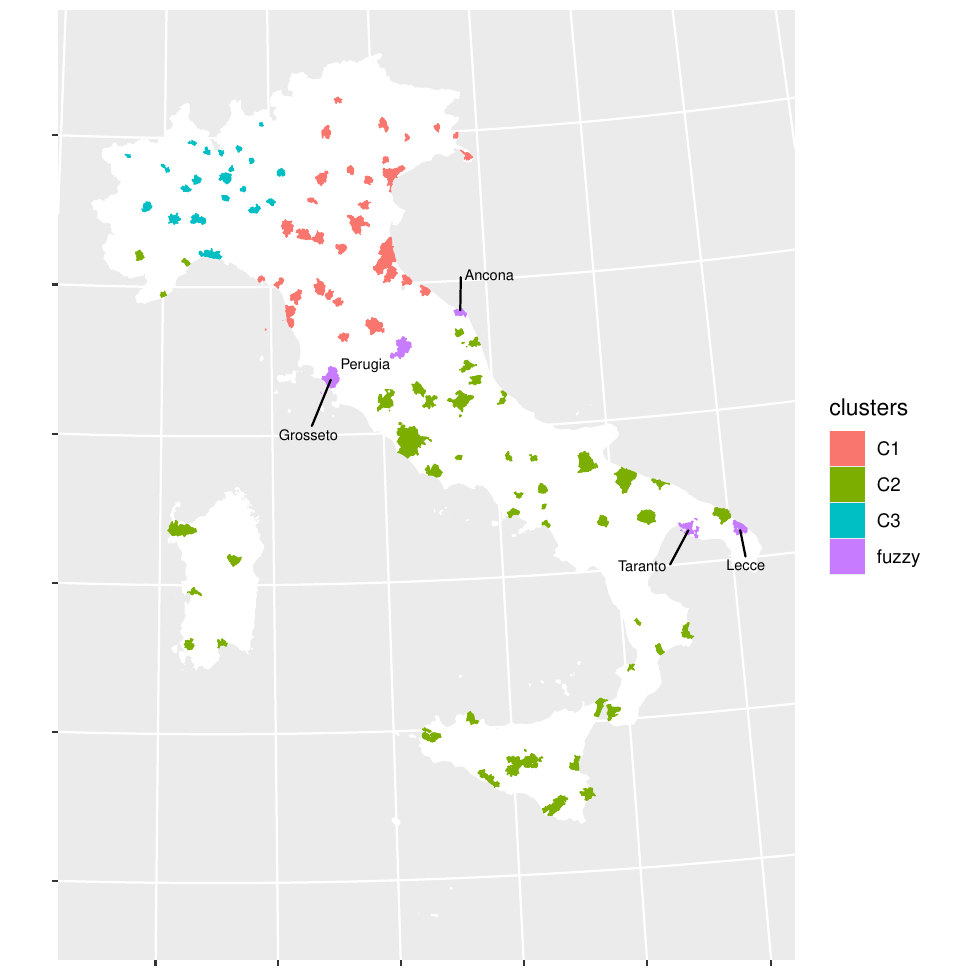}
    }
    \caption{Map of the crisp partitions based on  (a) $C=2$ and cutoff $0.7$ and (b) $C=3$ and cutoff $0.6$}
    \label{maps}
\end{figure}

\begin{figure}[!ht]
    \centering
    \subfloat[A1]{
        \includegraphics[width=6cm]{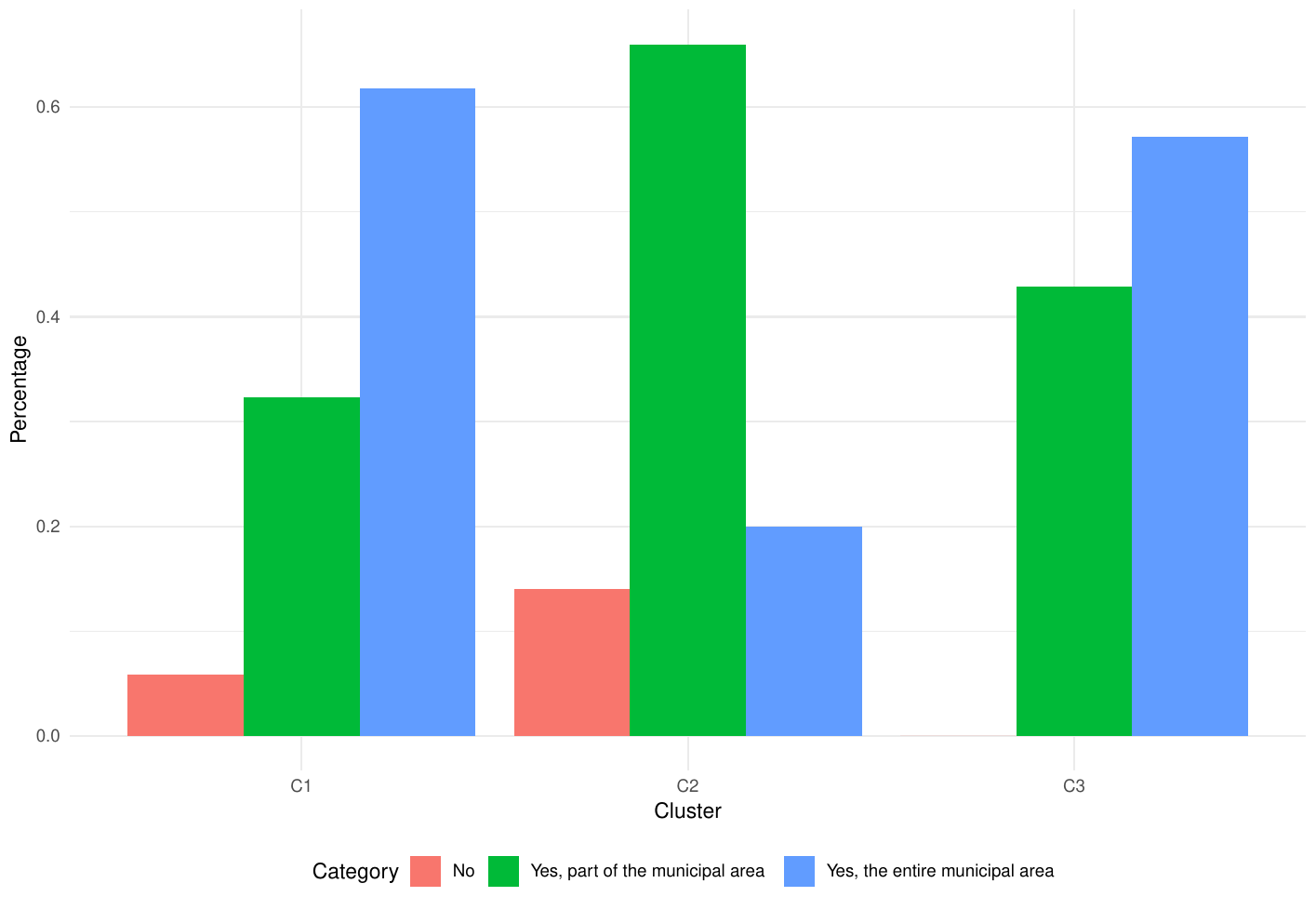}
    } 
    \subfloat[A2]{
        \includegraphics[width=6cm]{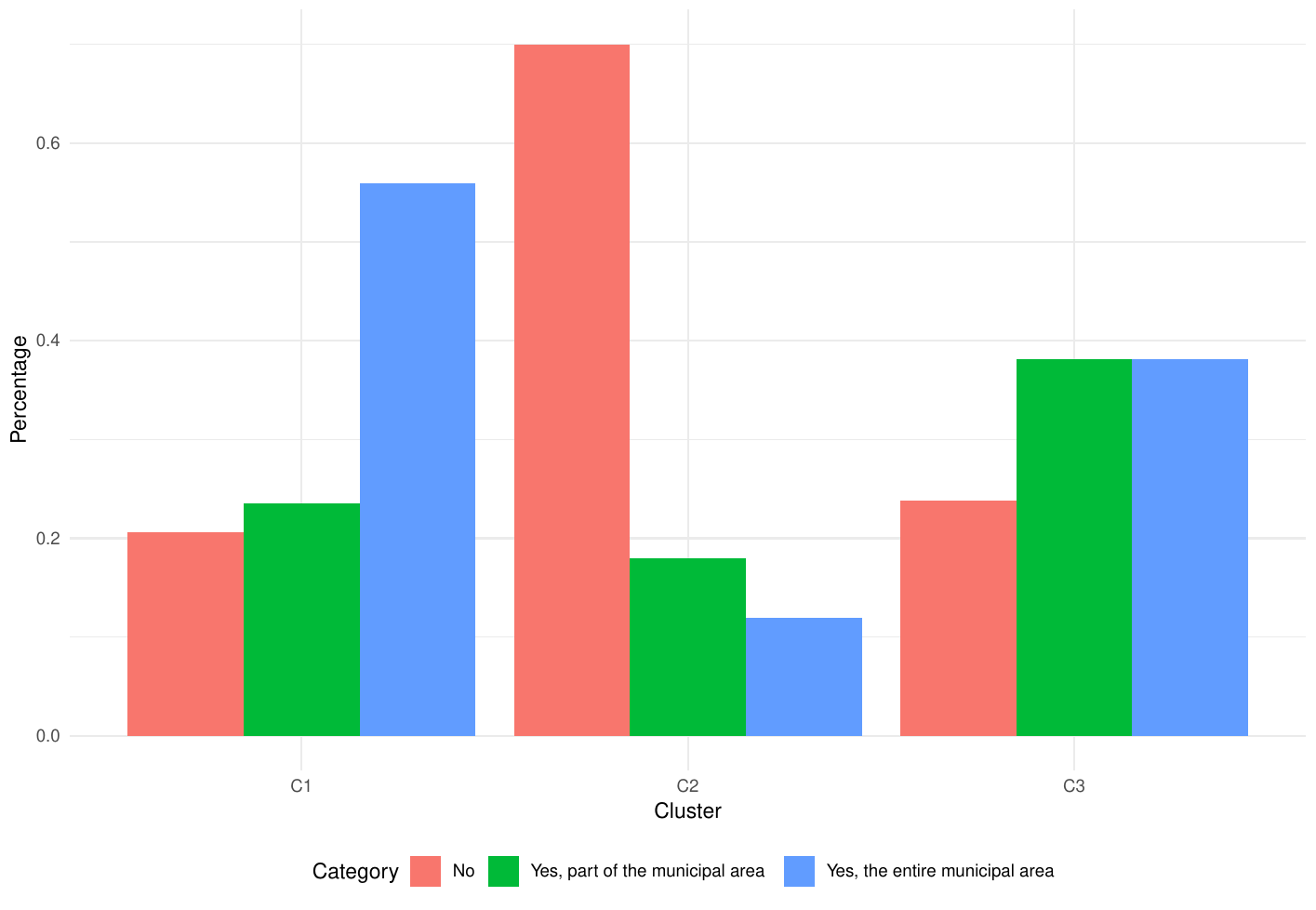}
    }\\
    \subfloat[A3]{
        \includegraphics[width=6cm]{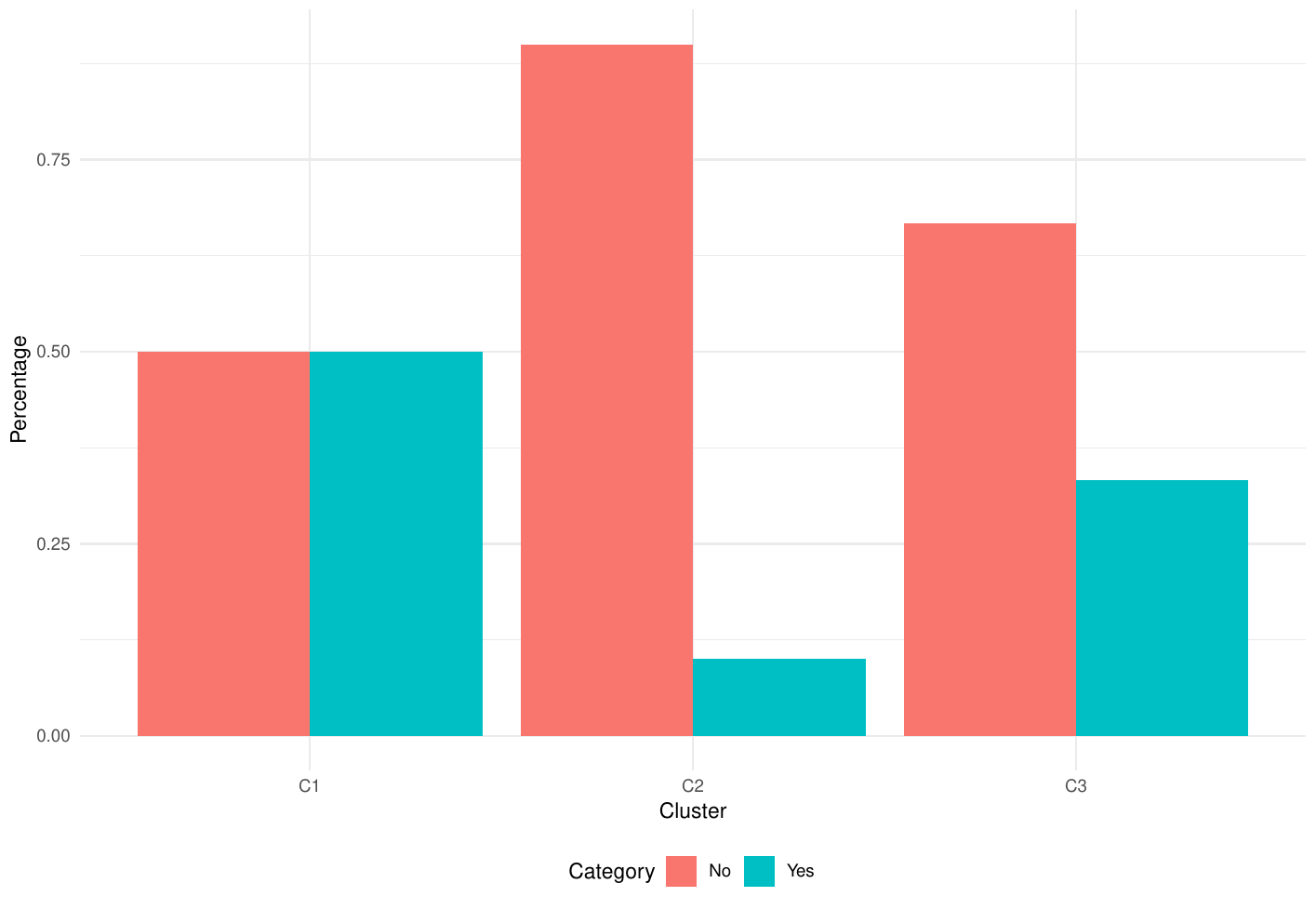}
    }
     \subfloat[A4]{
        \includegraphics[width=6cm]{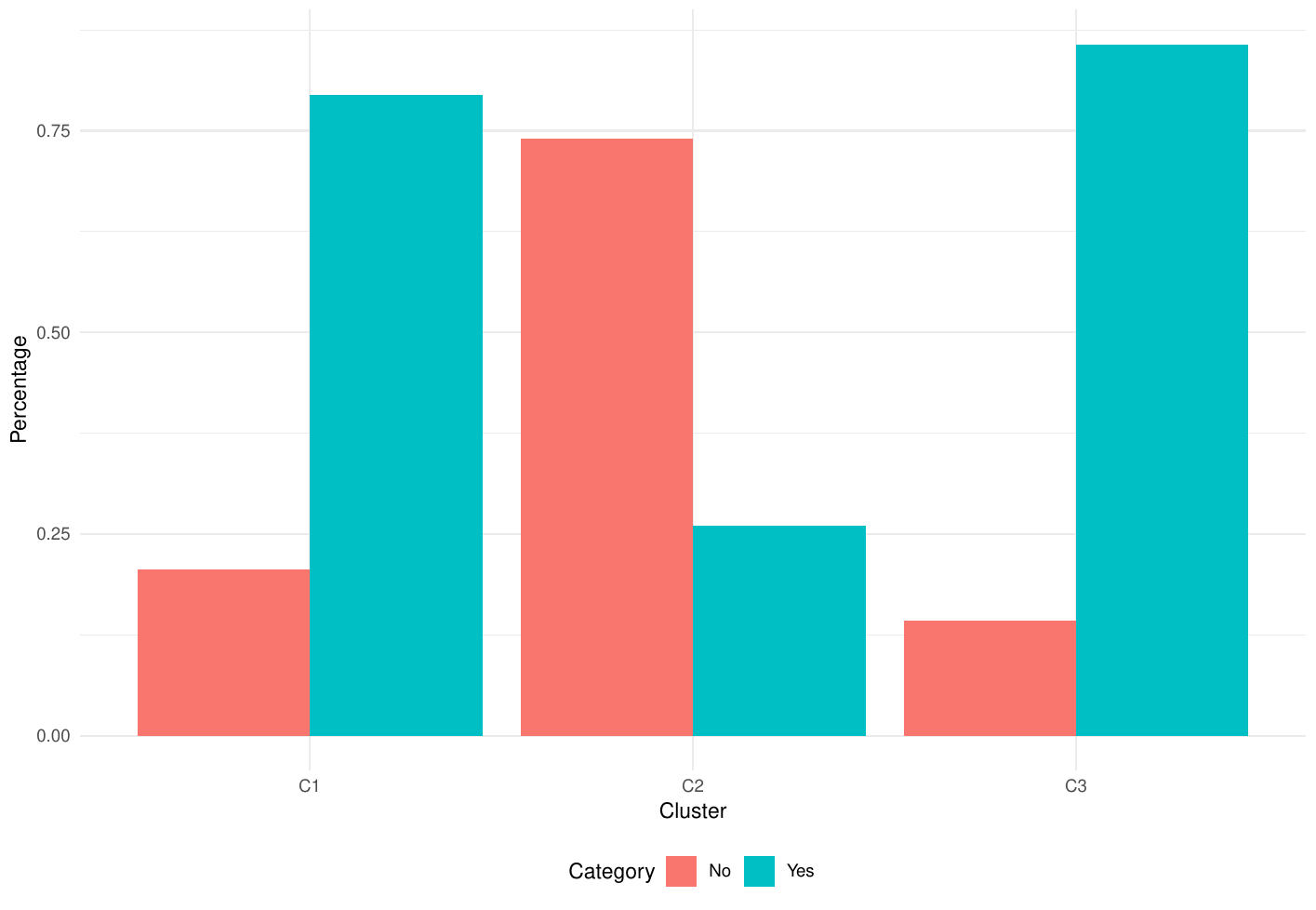}
    }\\
    \subfloat[A5]{
        \includegraphics[width=6cm]{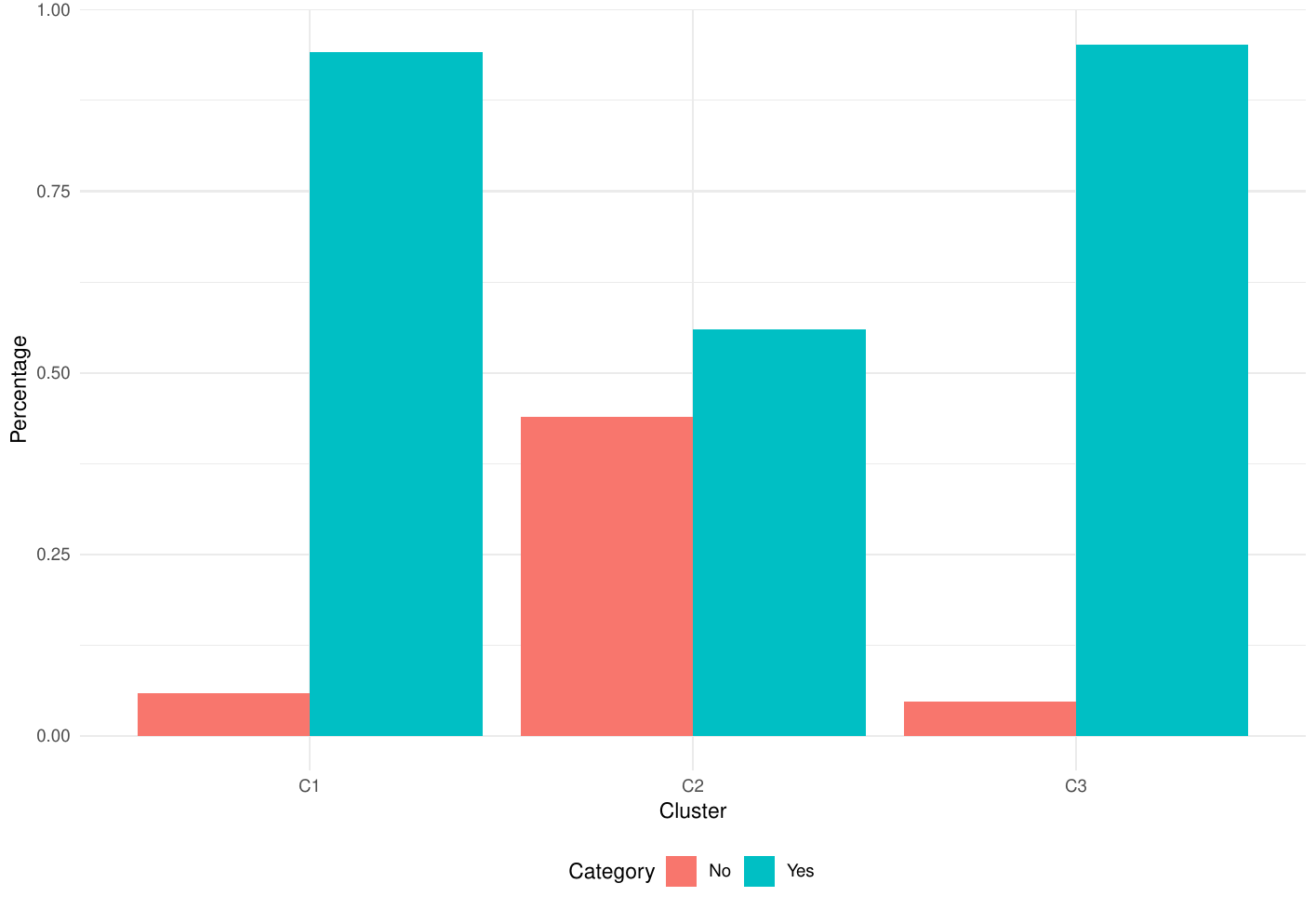}
    }
    \caption{barplots of variables A1-A5  according to clusters}
    \label{barplot1}
\end{figure}

\begin{figure}[!ht]
    \centering
    \subfloat[A6]{
        \includegraphics[width=6cm]{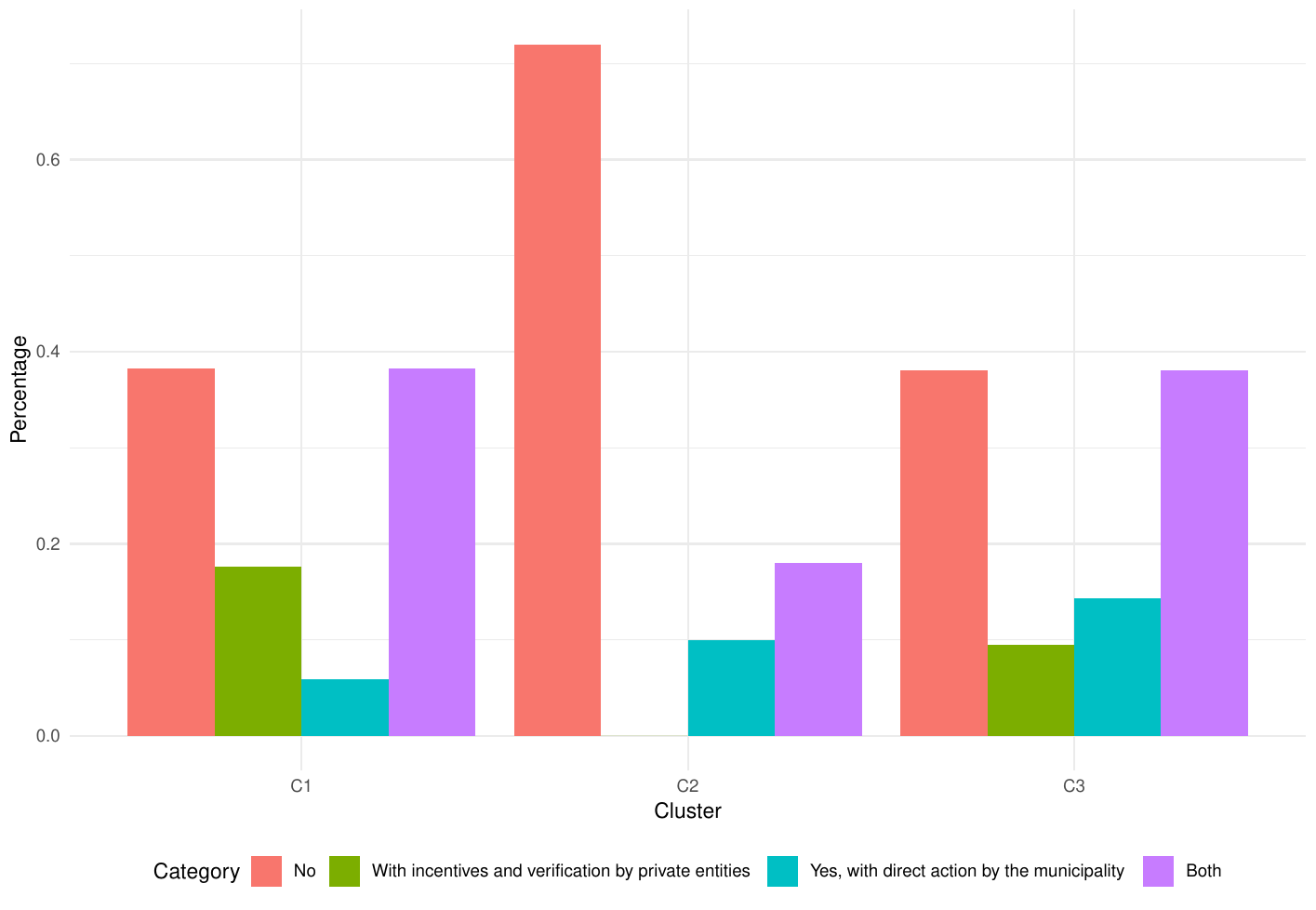}
    } 
    \subfloat[A7]{
        \includegraphics[width=6cm]{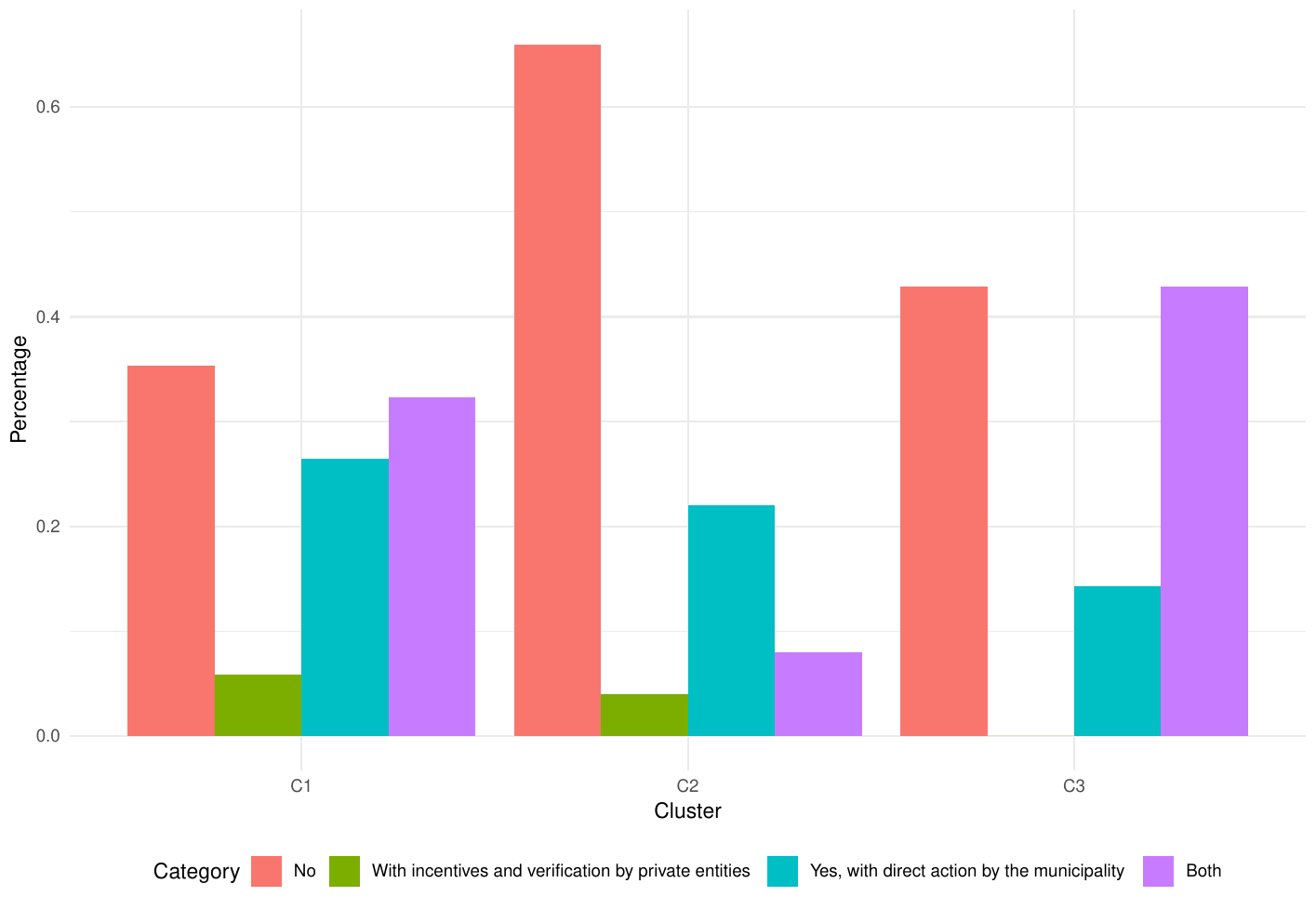}
    }\\
    \subfloat[A8]{
        \includegraphics[width=6cm]{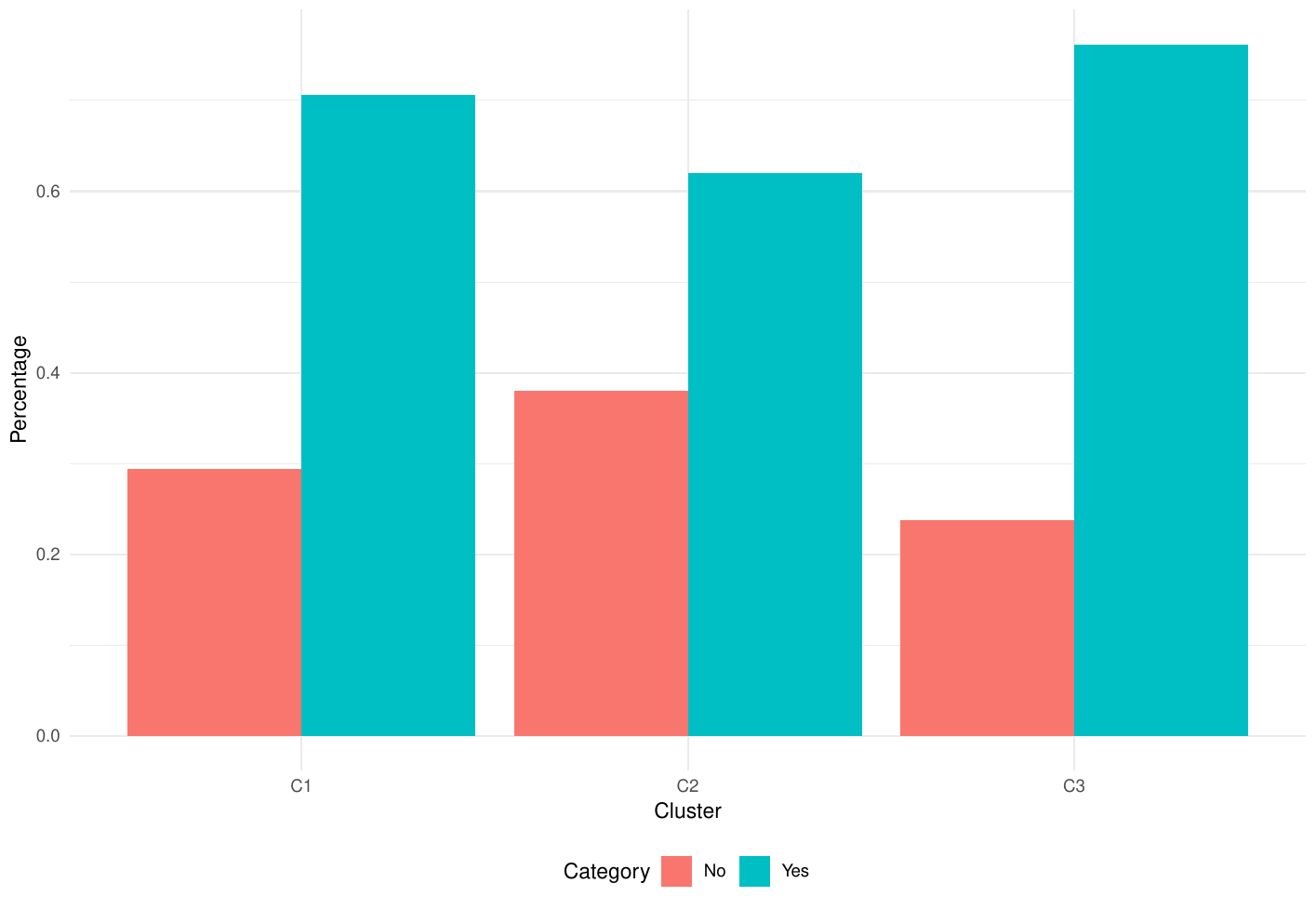}
    }
     \subfloat[A9]{
        \includegraphics[width=6cm]{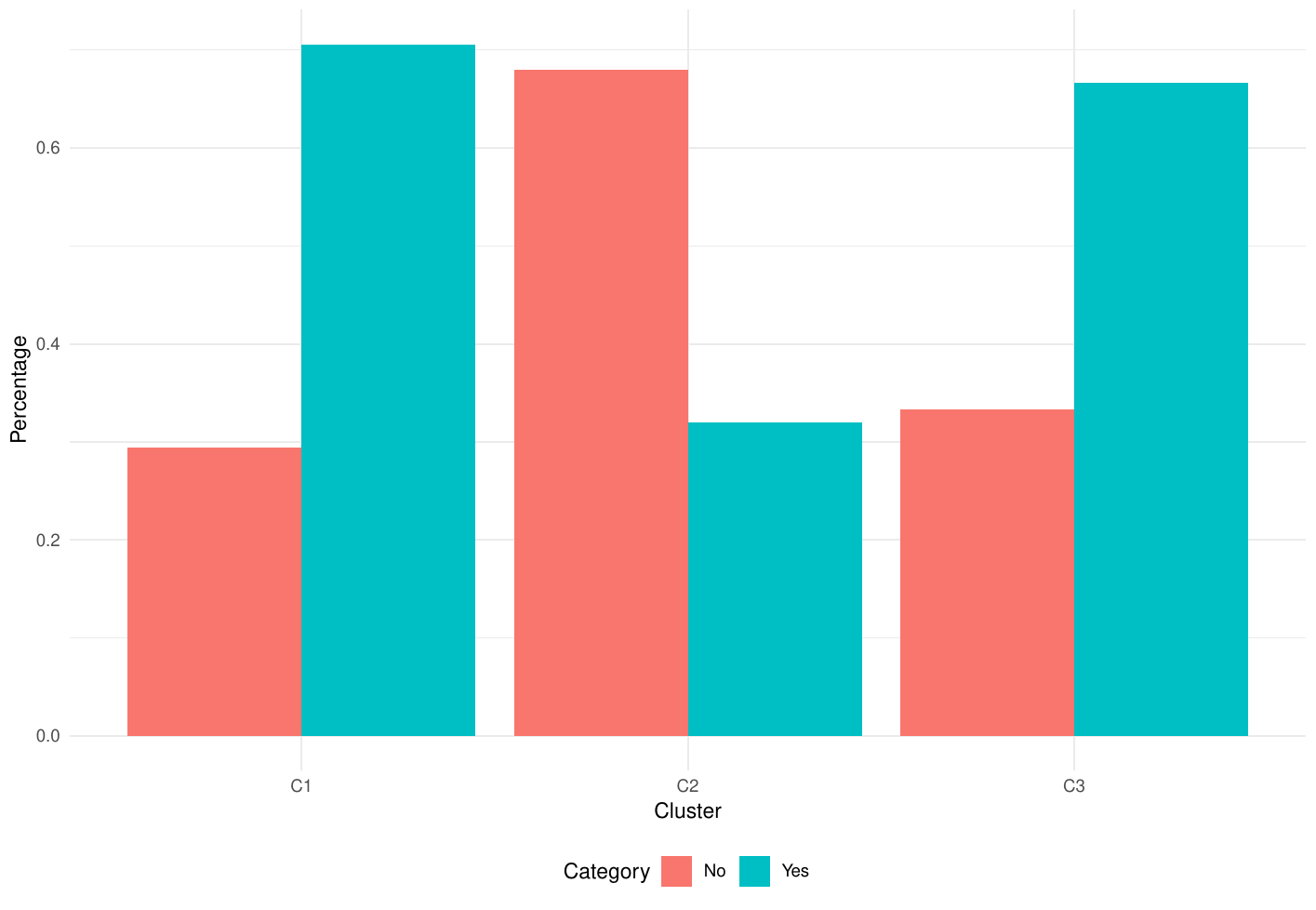}
    }
    \caption{Barplots of variables A6-A9  according to clusters.}
    \label{barplot2}
\end{figure}

The modes of the clusters are reported in Table \ref{tab:mode} for both solutions under consideration.
\begin{table}\resizebox{1\textwidth}{!}{
\begin{tabular}{l|p{2.2cm}p{2.2cm}|p{2.2cm}p{2.2cm}p{2.2cm}}
\toprule

        \multicolumn{1}{c}{$\gamma=0.9$}& \multicolumn{2}{c}{$C=2$} &\multicolumn{3}{c}{$C=3$} \\

 & C1 & C2 & C1 & C2 & C3 \\
\midrule
A1 & Yes, the entire municipal area  & Yes, part of the municipal area  & Yes, the entire municipal area  & Yes, part of the municipal area  & Yes, the entire municipal area \\
A2 & Yes, the entire municipal area  & no & Yes, the entire municipal area  & no & Yes, part of the municipal area  \\
A3 & no & no & no & no & no \\
A4 & yes & no & yes & no & yes \\
A5& yes & yes & yes & yes & yes \\
A6 & Both & no & no & no & no \\
A7 & Both & no & no & no & no \\
A8 & yes & yes & yes & yes & yes \\
A9 & yes & no & yes & no & yes\\
\bottomrule
\end{tabular}}
            \caption{The modes' values in the application of the FCMo-MSC $p=1$, $\gamma=0.9$ and $C=2,3$.}
            \label{tab:mode}
\end{table}
We present in Figure \ref{maps} the two maps of the crisp partitions, based on a membership cut-off of $0.7$ for $C=2$ and $0.6$ for $C=3$\footnote{Units are considered fuzzy if they have membership values in the ranges $(0.3, 0.7)$ for $C=2$ and if no membership value is larger than $0.6$ for $C=3$.}, to highlight the geographical distribution of the clusters. The matrix of the degrees of membership for both solutions is reported in Table \ref{partitions}.
As expected, the clustering results for both solutions reflect and enforce the geographical division between Northern and Southern Italy. Only a few of cities in the Northwest are classified within the cluster of Southern cities\footnote{For $C = 2$, the cities are Asti, Cuneo, Imperia, and Savona; for $C = 3$, the same cities are included but for Asti.}. The solution based on $C=3$ offers deeper insights into the differing behaviors of Northern regions regarding urban green space management and sustainability, revealing some interesting differences between the western and eastern parts. Therefore, we focused on interpreting the results derived from this analysis.
Based on barplots of Figs \ref{barplot1} and \ref{barplot2}, the following findings can be highlighted.

When examining the variables referred to census (A1 to A3), Cluster C1 stands out notably, as 61.76\% of cities in this cluster has completed the census across the entire municipal area. This reflects a strong commitment to environmental monitoring and management. However, 32.35\% of cities only has partial coverage, and 5.88\% has no census at all, indicating that while progress is being made, there remains room for improvement.

Cluster C2 presents a more concerning picture. The majority of cities in this cluster—66\%—has only partial coverage, and 14\% lacks any census data. This suggests that urban green management is not being prioritized, highlighting a significant gap in monitoring efforts compared to Cluster C1.

In contrast, Cluster C3 demonstrates a more balanced approach to urban green census. Here, 57.14\% of cities enjoys full coverage, while 42.86\% has only partial coverage. This indicates a varied level of commitment among cities, with some actively engaged in green space monitoring while others are still developing their strategies. Notably, there are no cities in Cluster C3 without an urban green census, which positively reflects overall engagement in green space management.

With respect to variable A2, cities in Clusters C1 and C3 significantly outperform those in Cluster C2. In Cluster C1, a notable 70.59\% of cities has integrated georeferenced data into their urban green census, and Cluster C3 follows with 66.67\%. Nevertheless, Cluster C1 stands out for its higher performance with 55.88\% of cities achieving full coverage, compared to only 38.10\% in Cluster C3. In contrast, cities in Cluster C2 show the opposite pattern, with 70\% having mostly no coverage and just 12\% attaining full coverage.

For variable A3, Cluster C1 particularly excels in the implementation of the Green Information System, though it is noteworthy that 50\% of cities has yet to adopt this system. This percentage rises significantly in Cluster C2, where 90\% of cities has not implemented the system, while in Cluster C3, the rate is 66.67\%.
Regarding variable A4, which refers to the commitment to transparency in urban green management, C1 and C3 show strong engagement. 79.41\% and 85.71\% of cities respectively publish their Tree Balance Sheets, which signifies a commitment to transparency and responsible urban tree management while C2 faces significant challenges, with 74.00\% of cities not publishing their Tree Balance Sheets.
For variable A5, Cluster C3 leads in monitoring efforts, with 95.24\% of cities actively assessing roadside tree failure risk, closely followed by Cluster C1 with 94.12\%. In contrast, Cluster C2 shows significantly lower engagement, with only 56\% of cities monitoring tree failure risks.

In Cluster C1, variable A6 indicates a moderate commitment to greening efforts, as evidenced by mixed results. Specifically, 38.24\% of cities has not implemented sufficient greening measures, while another 38.24\% employs a combination of municipal actions and private incentives for their greening initiatives. A smaller proportion (17.65\%) depends only on private entities for incentives and verification, and very few cities (5.88\%) engage in greening through direct municipal action alone. In Cluster C3, the results are similarly mixed, but with different types of actions: only 9.52\% of cities relies specfically on private incentives while 14.29\% are actively pursuing greening measures through direct municipal action alone.
Cluster C2, once again, faces substantial challenges, with 72.00\% of cities failing to adopt any greening measures and no engagement with private entities, highlighting a critical need for improvement in urban greening strategies.

The results for variable A7 indicate varying levels of commitment to conserving arboreal heritage among the clusters. Cluster C1 demonstrates a moderate commitment, with 26.47\% of municipalities actively implementing conservation measures. Additionally, 32.35\% utilizes a combination of municipal action and private incentives, while 5.88\% relies specifically on incentives and verification by private entities. Cluster C3 follows with percentages in these categories at 14.29\%, 42.86\%, and 0\%, respectively.

In contrast, Cluster C2 faces significant challenges in this area. Only a modest 22.00\% of cities is actively implementing conservation measures through direct municipal action. Furthermore,  8.00\% employs a dual approach, and very few cities (4.00\%) utilize private incentives, highlighting a critical need for enhanced policies and practices to promote arboreal heritage conservation in this cluster.

Results for variable A8 highlight that Clusters C1 and C3 demonstrate strong initiatives for community engagement.  70.59\% of cities in Cluster C1 and 76.19\% in Cluster C3 actively involve citizens or associations in the maintenance of urban green spaces. In contrast, this percentage decreases to 62.00\% for Cluster C2, indicating a need for improved engagement strategies in that cluster.

The data for the last variable A9 illustrates urban afforestation across the clusters, with Cluster C1 (70.59\%) leading in proactive measures, Cluster C3 (66.67\%) following closely, and Cluster C2  (32\%) requiring significant improvements.

In conclusion, Cluster C1, made up of northeastern provincial capitals, demonstrates a strong commitment to environmental management and citizen engagement. In contrast, Cluster C2, comprising central and southern provincial capitals, faces significant challenges, characterized by a high percentage of cities lacking adequate monitoring and transparency measures. The predominance of partial coverage and a concerning number of cities without any census data suggest a critical need for improvement in urban green management practices. This cluster's minimal engagement in greening efforts and tree safety monitoring highlights a significant gap in priorities compared to C1.
Cluster C3, made up of northwestern provincial capitals, presents a mixed picture, with a balanced approach to urban green management. While it exhibits strong engagement in certain areas, such as the absence of cities without census data, it still lags behind Cluster C1 in the implementation of robust green practices involving the entire municipal area.

Summing up, the contrasting levels of commitment to urban green management highlight the divide between the northern and southern cities in Italy, with the north-eastern cities, in particular, leading in proactive initiatives and comprehensive coverage, while the southern provinces struggle with significant gaps in monitoring and prioritization of environmental issues.


\section{Discussion and conclusions}

In this paper, we proposed using fuzzy modularity, as defined in \cite{nepusz2008fuzzy} as a spatial regularisation term in fuzzy clustering algorithms. We developed two algorithms, the Fuzzy C-Medoid with modularity spatial correction (FCMd-MSC) to be applied to numeric data, and the Fuzzy C-Modes with modularity spatial correction (FCMo-MSC), to be applied to categorical data. In both cases, the objective function to optimize is the convex combination of the two objective functions of a fuzzy clustering algorithm and a modularity maximisation algorithm plus an entropic term that tunes the fuzziness of the final partition. We showcased it, other than on simulated data, on two real-world datasets, one where spatiality represents proximity in a physical space and one where it describes abstract relations such as lack of travel restrictions. This follows the work done in \cite{d2024fuzzy} where we used the fuzzy Barber modularity as a spatial term in the joint clustering of two disjoint sets of units, linked by a bipartite adjacency structure.

Our approach is meant to build clusters that are as much as possible both valid clusters in terms of the attributes of the units and strong communities in terms of the adjacency network. In our model, a priori the information carried by the attributes and by the adjacency matrix are given separately and have the same importance. We let the parameter $\gamma$ determine their relative weight. The optimal value of $\gamma$ can be optimised using an appropriate validity function or set by the user to indicate the subjective relevance of the two types of information for the classification problem at hand. We have shown in the simulations in Section \ref{sec:sim} that indeed when the natural clusters based only on the attributes and only on the network structure are different, Algorithms \ref{alg:md} and \ref{alg:mo}, based on the values of $\gamma$ and the number of clusters $C$ can either find one of the two configurations or combine them into a new cluster structure that takes into account both. This is unlike other methods for spatial clustering such as the penalty function introduced in \cite{Pham2001}, where the adjacency matrix only acts as a correction term. In such models, increasing the spatial term causes the output to degenerate into a trivial partition in which all units are in the same cluster.

We have shown that the approach based on optimizing a convex combination of modularity and an objective function of fuzzy clustering is effective when applied to the Fuzzy C-Medoids with Euclidean distance and the Fuzzy C-Modes. We believe it would be worth investigating its applicability to other spatial clustering problems, as we do not foresee serious obstacles to its adaptation.

\noindent
\begin{appendices}

\section{Membership tables for applications to real data}
\label{app:longtables}
\begin{longtable}{p{5cm}|l l|p{3cm}}
    \toprule
           Countries & Cluster 1 & Cluster 2& Crisp Partition \\ 
        \midrule
Afghanistan     	&	0.86	&	0.14	&	1	\\
Angola  	&	0.96	&	0.04	&	1	\\
Bangladesh      	&	0.83	&	0.17	&	1	\\
Belarus 	&	0.78	&	0.22	&	1	\\
Bolivia 	&	0.98	&	0.02	&	1	\\
Brazil  	&	0.96	&	0.04	&	1	\\
Cambodia        	&	0.86	&	0.14	&	1	\\
Cameroon        	&	0.94	&	0.06	&	1	\\
Colombia        	&	0.89	&	0.11	&	1	\\
Dominican Republic      	&	0.99	&	0.01	&	1	\\
Egypt, Arab Rep.        	&	0.71	&	0.29	&	1	\\
Ghana   	&	0.97	&	0.03	&	1	\\
\textbf{Guyana} 	&	       \textbf{1.00}   	&	       \textbf{0.00}   	&	1	\\
Haiti   	&	0.96	&	0.04	&	1	\\
Honduras        	&	0.84	&	0.16	&	1	\\
Jamaica 	&	1.00	&	0.00	&	1	\\
Liberia 	&	0.98	&	0.02	&	1	\\
Malawi  	&	0.96	&	0.04	&	1	\\
Mexico  	&	0.79	&	0.21	&	1	\\
Mongolia        	&	0.95	&	0.05	&	1	\\
Myanmar 	&	0.94	&	0.06	&	1	\\
Niger   	&	0.96	&	0.04	&	1	\\
Nigeria 	&	0.98	&	0.02	&	1	\\
Pakistan        	&	0.85	&	0.15	&	1	\\
Panama  	&	0.83	&	0.17	&	1	\\
Paraguay        	&	0.86	&	0.14	&	1	\\
Peru    	&	0.93	&	0.07	&	1	\\
Sierra Leone    	&	0.96	&	0.04	&	1	\\
South Africa    	&	1.00	&	0.00	&	1	\\
Sri Lanka       	&	0.83	&	0.17	&	1	\\
Suriname        	&	0.91	&	0.09	&	1	\\
Tanzania        	&	0.98	&	0.02	&	1	\\
Trinidad and Tobago     	&	0.96	&	0.04	&	1	\\
Tunisia 	&	0.89	&	0.11	&	1	\\
Türkiye 	&	0.78	&	0.22	&	1	\\
Uganda  	&	0.97	&	0.03	&	1	\\
Vietnam 	&	0.78	&	0.22	&	1	\\
Zambia  	&	0.98	&	0.02	&	1	\\
Zimbabwe        	&	0.99	&	0.01	&	1	\\
Albania 	&	0.23	&	0.77	&	2	\\
Australia       	&	0.05	&	0.95	&	2	\\
Austria 	&	0.01	&	0.99	&	2	\\
Belgium 	&	0.02	&	0.98	&	2	\\
Canada  	&	0.06	&	0.94	&	2	\\
Czechia 	&	0.02	&	0.98	&	2	\\
Denmark 	&	0.01	&	0.99	&	2	\\
Estonia 	&	0.01	&	0.99	&	2	\\
Finland      	&	0.01	&	0.99	&	2	\\
France  	&	0.06	&	0.94	&	2	\\
Georgia 	&	0.06	&	0.94	&	2	\\
Germany 	&	0.02	&	0.98	&	2	\\
Greece  	&	0.13	&	0.87	&	2	\\
Hungary 	&	0.15	&	0.85	&	2	\\
Ireland 	&	0.01	&	0.99	&	2	\\
Italy   	&	0.08	&	0.92	&	2	\\
Japan   	&	0.01	&	0.99	&	2	\\
\textbf{Latvia} 	&	       \textbf{0.00}   	&	       \textbf{1.00}   	&	2	\\
Lithuania       	&	0.02	&	0.98	&	2	\\
Luxembourg      	&	0.01	&	0.99	&	2	\\
Netherlands     	&	0.03	&	0.97	&	2	\\
New Zealand     	&	0.06	&	0.94	&	2	\\
North Macedonia 	&	0.06	&	0.94	&	2	\\
Norway  	&	0.01	&	0.99	&	2	\\
Poland  	&	0.05	&	0.95	&	2	\\
Portugal        	&	0.04	&	0.96	&	2	\\
Serbia  	&	0.24	&	0.76	&	2	\\
Slovak Republic 	&	0.04	&	0.96	&	2	\\
Slovenia        	&	0.02	&	0.98	&	2	\\
Spain   	&	0.04	&	0.96	&	2	\\
Sweden  	&	0.02	&	0.98	&	2	\\
Ukraine 	&	0.23	&	0.77	&	2	\\
United States   	&	0.27	&	0.73	&	2	\\
Algeria 	&	0.66	&	0.34	&	Fuzzy	\\
Chile   	&	0.57	&	0.43	&	Fuzzy	\\
Costa Rica      	&	0.32	&	0.68	&	Fuzzy	\\
El Salvador     	&	0.45	&	0.55	&	Fuzzy	\\
Jordan  	&	0.48	&	0.52	&	Fuzzy	\\
Uruguay  \footnote{Uruguay is included as a fuzzy unit, as its membership degree in the second cluster is rounded up from 0.697}	&	0.30	&	0.70	&	Fuzzy	\\
         \bottomrule
            \caption{The $U$ matrix in the application of the FCMd-MSC with $p=3$, $\gamma=0.7$ and $C=2$.}
            \label{partitions0}
\end{longtable}

    \begin{longtable}{l|l l|ll l}
    \toprule
        \multicolumn{1}{c}{$\gamma=0.9$}& \multicolumn{2}{c}{$C=2$} &\multicolumn{3}{c}{$C=3$} \\
        COMUNI & C1 & C2 & C1 & C2 & C3 \\ \hline
        \midrule

        Agrigento & 0.00 & 1.00 & 0.02 & 0.96 & 0.02 \\ 
        Alessandria & 0.98 & 0.02 & 0.00 & 0.00 & 1.00 \\ 
        Ancona & 0.02 & 0.98 & 0.49 & 0.39 & 0.12 \\ 
        Andria & 0.00 & 1.00 & 0.00 & 1.00 & 0.00 \\ 
        Aosta & 0.48 & 0.52 & 0.02 & 0.03 & 0.95 \\ 
        Arezzo & 0.99 & 0.01 & 1.00 & 0.00 & 0.00 \\ 
        Ascoli Piceno & 0.00 & 1.00 & 0.00 & 1.00 & 0.00 \\ 
        Asti & 0.04 & 0.96 & 0.00 & 0.01 & 0.99 \\ 
        Avellino & 0.00 & 1.00 & 0.00 & 0.99 & 0.00 \\ 
        Bari & 0.00 & 1.00 & 0.01 & 0.97 & 0.01 \\ 
        Barletta & 0.00 & 1.00 & 0.00 & 1.00 & 0.00 \\ 
        Belluno & 0.97 & 0.03 & 1.00 & 0.00 & 0.00 \\ 
        Benevento & 0.00 & 1.00 & 0.00 & 1.00 & 0.00 \\ 
        Bergamo & 1.00 & 0.00 & 0.00 & 0.00 & 1.00 \\ 
        Biella & 0.96 & 0.04 & 0.00 & 0.00 & 1.00 \\ 
        Bologna & 1.00 & 0.00 & 1.00 & 0.00 & 0.00 \\ 
        Bolzano & 0.96 & 0.04 & 0.80 & 0.16 & 0.04 \\ 
        Brescia & 1.00 & 0.00 & 0.01 & 0.00 & 0.99 \\ 
        Brindisi & 0.00 & 1.00 & 0.08 & 0.81 & 0.11 \\ 
        Cagliari & 0.00 & 1.00 & 0.06 & 0.78 & 0.16 \\ 
        Caltanissetta & 0.00 & 1.00 & 0.00 & 1.00 & 0.00 \\ 
        Campobasso & 0.00 & 1.00 & 0.00 & 1.00 & 0.00 \\ 
        Carbonia & 0.00 & 1.00 & 0.07 & 0.85 & 0.08 \\ 
        Caserta & 0.00 & 1.00 & 0.00 & 1.00 & 0.00 \\ 
        Catania & 0.00 & 1.00 & 0.00 & 1.00 & 0.00 \\ 
        Catanzaro & 0.03 & 0.97 & 0.15 & 0.75 & 0.11 \\ 
        Cesena & 0.99 & 0.01 & 1.00 & 0.00 & 0.00 \\ 
        Chieti & 0.00 & 1.00 & 0.00 & 1.00 & 0.00 \\ 
        Como & 1.00 & 0.00 & 0.00 & 0.00 & 1.00 \\ 
        Cosenza & 0.00 & 1.00 & 0.01 & 0.98 & 0.01 \\ 
        Cremona & 1.00 & 0.00 & 0.01 & 0.00 & 0.99 \\ 
        Crotone & 0.00 & 1.00 & 0.01 & 0.98 & 0.01 \\ 
        Cuneo & 0.02 & 0.98 & 0.02 & 0.85 & 0.13 \\ 
        Enna & 0.00 & 1.00 & 0.00 & 1.00 & 0.00 \\ 
        Fermo & 0.00 & 1.00 & 0.00 & 0.99 & 0.00 \\ 
        Ferrara & 1.00 & 0.00 & 1.00 & 0.00 & 0.00 \\ 
        Firenze & 1.00 & 0.00 & 1.00 & 0.00 & 0.00 \\ 
        Foggia & 0.00 & 1.00 & 0.00 & 0.99 & 0.00 \\ 
        Forlì & 1.00 & 0.00 & 1.00 & 0.00 & 0.00 \\ 
        Frosinone & 0.00 & 1.00 & 0.02 & 0.96 & 0.03 \\ 
        Genova & 0.72 & 0.28 & 0.09 & 0.04 & 0.87 \\ 
        Gorizia & 0.73 & 0.27 & 0.92 & 0.03 & 0.05 \\ 
        Grosseto & 0.57 & 0.43 & 0.40 & 0.15 & 0.45 \\ 
        Imperia & 0.00 & 1.00 & 0.01 & 0.97 & 0.02 \\ 
        Isernia & 0.00 & 1.00 & 0.00 & 1.00 & 0.00 \\ 
        La Spezia & 0.93 & 0.07 & 0.99 & 0.00 & 0.00 \\ 
        L'Aquila & 0.00 & 1.00 & 0.00 & 1.00 & 0.00 \\ 
        Latina & 0.04 & 0.96 & 0.13 & 0.82 & 0.06 \\ 
        Lecce & 0.09 & 0.91 & 0.19 & 0.55 & 0.25 \\ 
        Lecco & 1.00 & 0.00 & 0.00 & 0.00 & 1.00 \\ 
        Livorno & 1.00 & 0.00 & 1.00 & 0.00 & 0.00 \\ 
        Lodi & 1.00 & 0.00 & 0.00 & 0.00 & 1.00 \\ 
        Lucca & 1.00 & 0.00 & 1.00 & 0.00 & 0.00 \\ 
        Macerata & 0.00 & 1.00 & 0.04 & 0.95 & 0.01 \\ 
        Mantova & 1.00 & 0.00 & 0.98 & 0.00 & 0.02 \\ 
        Massa & 0.99 & 0.01 & 0.99 & 0.01 & 0.00 \\ 
        Matera & 0.00 & 1.00 & 0.00 & 1.00 & 0.00 \\ 
        Messina & 0.00 & 1.00 & 0.01 & 0.96 & 0.03 \\ 
        Milano & 1.00 & 0.00 & 0.00 & 0.00 & 1.00 \\ 
        Modena & 1.00 & 0.00 & 1.00 & 0.00 & 0.00 \\ 
        Monza & 1.00 & 0.00 & 0.00 & 0.00 & 1.00 \\ 
        Napoli & 0.00 & 1.00 & 0.00 & 0.99 & 0.00 \\ 
        Novara & 1.00 & 0.00 & 0.00 & 0.00 & 1.00 \\ 
        Nuoro & 0.01 & 0.99 & 0.08 & 0.83 & 0.09 \\ 
        Oristano & 0.00 & 1.00 & 0.01 & 0.97 & 0.02 \\ 
        Padova & 1.00 & 0.00 & 1.00 & 0.00 & 0.00 \\ 
        Palermo & 0.03 & 0.97 & 0.09 & 0.73 & 0.18 \\ 
        Parma & 0.99 & 0.01 & 0.94 & 0.00 & 0.06 \\ 
        Pavia & 1.00 & 0.00 & 0.00 & 0.00 & 1.00 \\ 
        Perugia & 0.12 & 0.88 & 0.39 & 0.48 & 0.13 \\ 
        Pesaro & 0.23 & 0.77 & 0.89 & 0.10 & 0.01 \\ 
        Pescara & 0.00 & 1.00 & 0.00 & 0.99 & 0.00 \\ 
        Piacenza & 1.00 & 0.00 & 0.00 & 0.00 & 1.00 \\ 
        Pisa & 0.98 & 0.02 & 1.00 & 0.00 & 0.00 \\ 
        Pistoia & 1.00 & 0.00 & 1.00 & 0.00 & 0.00 \\ 
        Pordenone & 1.00 & 0.00 & 1.00 & 0.00 & 0.00 \\ 
        Potenza & 0.00 & 1.00 & 0.00 & 1.00 & 0.00 \\ 
        Prato & 1.00 & 0.00 & 1.00 & 0.00 & 0.00 \\ 
        Ragusa & 0.00 & 1.00 & 0.00 & 1.00 & 0.00 \\ 
        Ravenna & 0.95 & 0.05 & 1.00 & 0.00 & 0.00 \\ 
        Reggio di Calabria & 0.00 & 1.00 & 0.00 & 0.99 & 0.00 \\ 
        Reggio nell'Emilia & 1.00 & 0.00 & 1.00 & 0.00 & 0.00 \\ 
        Rieti & 0.00 & 1.00 & 0.00 & 1.00 & 0.00 \\ 
        Rimini & 0.94 & 0.06 & 0.97 & 0.01 & 0.02 \\ 
        Roma & 0.00 & 1.00 & 0.00 & 0.99 & 0.01 \\ 
        Rovigo & 1.00 & 0.00 & 1.00 & 0.00 & 0.00 \\ 
        Salerno & 0.00 & 1.00 & 0.00 & 1.00 & 0.00 \\ 
        Sassari & 0.01 & 0.99 & 0.13 & 0.71 & 0.15 \\ 
        Savona & 0.00 & 1.00 & 0.01 & 0.90 & 0.09 \\ 
        Siena & 0.85 & 0.15 & 1.00 & 0.00 & 0.00 \\ 
        Siracusa & 0.00 & 1.00 & 0.02 & 0.92 & 0.06 \\ 
        Sondrio & 1.00 & 0.00 & 0.00 & 0.00 & 1.00 \\ 
        Taranto & 0.03 & 0.97 & 0.19 & 0.47 & 0.34 \\ 
        Teramo & 0.00 & 1.00 & 0.00 & 1.00 & 0.00 \\ 
        Terni & 0.00 & 1.00 & 0.00 & 1.00 & 0.00 \\ 
        Torino & 0.48 & 0.52 & 0.00 & 0.00 & 0.99 \\ 
        Trani & 0.00 & 1.00 & 0.00 & 1.00 & 0.00 \\ 
        Trapani & 0.00 & 1.00 & 0.03 & 0.93 & 0.04 \\ 
        Trento & 1.00 & 0.00 & 0.97 & 0.00 & 0.03 \\ 
        Treviso & 1.00 & 0.00 & 1.00 & 0.00 & 0.00 \\ 
        Trieste & 0.54 & 0.46 & 0.85 & 0.08 & 0.08 \\ 
        Udine & 0.97 & 0.03 & 0.99 & 0.00 & 0.00 \\ 
        Varese & 1.00 & 0.00 & 0.00 & 0.00 & 1.00 \\ 
        Venezia & 1.00 & 0.00 & 1.00 & 0.00 & 0.00 \\ 
        Verbania & 1.00 & 0.00 & 0.00 & 0.00 & 1.00 \\ 
        Vercelli & 0.99 & 0.01 & 0.00 & 0.00 & 1.00 \\ 
        Verona & 1.00 & 0.00 & 1.00 & 0.00 & 0.00 \\ 
        Vibo Valentia & 0.00 & 1.00 & 0.01 & 0.98 & 0.02 \\ 
        Vicenza & 1.00 & 0.00 & 1.00 & 0.00 & 0.00 \\ 
        Viterbo & 0.00 & 1.00 & 0.00 & 1.00 & 0.00 \\ 
        \bottomrule
            \caption{The units' membership values in the application of the FCMo-MSC $p=1$, $\gamma=0.9$ and $C=2,3$. \label{partitions}}
\end{longtable}

\end{appendices}

\section*{Declarations}

\begin{itemize}
\item Funding: This research received no specific funding.
\item Conflict of Interest: The authors have no conflict of interest to declare
\item Ethical Conduct: The paper does not include any experimental study with humans or animals.
\item Data Availability Statements: The data and code used in the simulations and application is available from the corresponding author upon reasonable request.
\end{itemize}

\bibliography{sn-bibliography}

\end{document}